\documentclass[aps,preprint,showpacs,preprintnumbers,amsmath,amssymb,floatfix]{revtex4}
\usepackage{graphicx}
\usepackage{amssymb,amsmath}

\def\beq{\begin{equation}}
\def\eeq{\end{equation}}
\def\beqn{\begin{eqnarray}}
\def\eeqn{\end{eqnarray}}

\def\r {{\bf r}}
\def\x {{\bf x}}
\def\y {{\bf y}}
\def\z {{\bf z}}
\def\n {{\bf n}}
\def\i {{\bf i}}
\def\brho {\mbox{\boldmath $\rho$}}

\begin{document} 
%%\begin{frontmatter}

\title{Recursive formulation of the 
multiconfigurational time-dependent Hartree method
for fermions, bosons and mixtures thereof  
in terms of one-body density operators}

\author{Ofir E. Alon$^{1\ast}$\footnote[0]{$^{\ast}$ ofir@research.haifa.ac.il},
Alexej I. Streltsov$^{2\dag}$\footnote[0]{$^{\dag}$ alexej.streltsov@pci.uni-heidelberg.de},
Kaspar Sakmann$^{2\ddag}$\footnote[0]{$^{\ddag}$ kaspar.sakmann@pci.uni-heidelberg.de},\break
Axel U. J. Lode$^{2\S}$\footnote[0]{$^{\S}$ axel.lode@pci.uni-heidelberg.de},
Julian Grond$^{2\P}$\footnote[0]{$^{\P}$ julian.grond@pci.uni-heidelberg.de},
and Lorenz S. Cederbaum$^{2\parallel}$\footnote[0]{$^{\parallel}$ lorenz.cederbaum@pci.uni-heidelberg.de}}

\affiliation{$^{1}$ Department of Physics, University of Haifa at Oranim, Tivon 36006, Israel.}

\affiliation{$^{2}$ Theoretische Chemie, Physikalisch-Chemisches Institut, Universit\"at Heidelberg,\\
Im Neuenheimer Feld 229, D-69120 Heidelberg, Germany.}

%% \date{\today}

%% \newpage

\begin{abstract}
The multiconfigurational time-dependent Hartree method (MCTDH)
[H.-D. Meyer, U. Manthe, and L. S. Cederbaum, Chem. Phys. Lett. {\bf 165},
73 (1990); U. Manthe, H.-D. Meyer, and L. S. Cederbaum, J. Chem. Phys. {\bf 97}, 3199
(1992)] is celebrating nowadays entering its third decade of 
tackling numerically-exactly a broad range of correlated 
multi-dimensional non-equilibrium quantum dynamical systems.
Taking in recent years particles' statistics explicitly
into account,
within the MCTDH for fermions (MCTDHF) and for bosons (MCTDHB),
has opened up further opportunities to treat larger systems 
of interacting identical particles, 
primarily in laser-atom and cold-atom physics.
With the increase of experimental capabilities to
simultaneously trap mixtures of two, three, and possibly even 
multiple kinds of
interacting composite identical particles together,
we set up the stage in the present work and 
specify the MCTDH method for such cases.
Explicitly,
the MCTDH method for systems with three kinds of identical
particles interacting via all combinations of two- and three-body forces is presented, 
and the resulting equations-of-motion are briefly discussed.
All four possible mixtures (Fermi-Fermi-Fermi,
Bose-Fermi-Fermi, Bose-Bose-Fermi and Bose-Bose-Bose) 
are presented in a unified manner.
Particular attention is paid to represent the 
coefficients' part of the equations-of-motion
in a compact recursive form 
in terms of one-body density operators only.
The recursion utilizes the recently proposed 
Combinadic-based mapping for fermionic and bosonic operators in Fock space
[A. I. Streltsov, O. E. Alon, and L. S. Cederbaum, Phys. Rev. A {\bf 81}, 022124 (2010)] 
and successfully applied and implemented within MCTDHB.
Our work sheds new light on the
representation of the coefficients'
part in MCTDHF and MCTDHB 
without resorting to the 
matrix elements of the many-body Hamiltonian 
with respect to the time-dependent configurations.
It suggests a recipe for 
efficient implementation of 
the schemes derived here for mixtures
which is suitable for parallelization.
\end{abstract}

%%\begin{keyword} Time-dependent many-particle Schr\"odinger equation;
%%Dirac-Frenkel variational principle;
%%Multiconfigurational time-dependent Hartree (MCTDH); 
%%Reduced density matrices;
%%Fock-space mapping of bosonic and fermionic operators;
%%MCTDH for fermions (MCTDHF); 
%%MCTDH for bosons (MCTDHB); 
%%MCTDH for Fermi-Fermi-Fermi mixtures (MCTDH-FFF); 
%%MCTDH for Bose-Fermi-Fermi mixtures (MCTDH-BFF); 
%%MCTDH for Bose-Bose-Fermi mixtures (MCTDH-BBF); 
%%MCTDH for Bose-Bose-Bose mixtures (MCTDH-BBB).
%%\end{keyword}

\pacs{31.15.xv, 67.60.-g, 05.30.Fk, 05.30.Jp, 03.65.-w}

\maketitle

%%\end{frontmatter}

\section{Introduction}\label{SEC1}

Quantum non-equilibrium dynamics is important to many branches of physics and chemistry 
\cite{Book_dynamics1,Book_dynamics2,Nuclear_book,Book_dynamics3,Pit_Stri_book,Book_dynamics4}
and often requires the solution of the time-dependent 
many-particle Schr\"odinger equation.
A particular efficient method
to solve the time-dependent
many-particle Schr\"odinger equation
is the multiconfigurational time-dependent Hartree
(MCTDH) algorithm and approach \cite{cpl,jcp,review,book}.
MCTDH,
which is considered at present the most efficient wave-packet
propagation tool,
has amply been employed for multi-dimensional
dynamical systems of distinguishable degrees-of-freedom,
typically molecular vibrations, see, e.g., 
Refs.~\cite{JCP_24a,JCP_24b,Manthe_review,Lenz_CI,relaxation2,vib_new1,vib_new2,irene}.
We mention that recent developments on multi-layer formulation of MCTDH
have opened up further possibilities to treat 
larger systems of distinguishable 
degrees-of-freedom \cite{ML_1,ML_2,ML_3}.
MCTDH has recently been applied with much success to various 
systems with a few identical particles
in the field of cold-atom physics,
see, e.g., Refs.~\cite{ZO_st1,ZO_st2,ZO_dy2,Sascha_mix,axel,Sascha_dip}.

In recent years, 
taking the quantum statistics between identical particles {\it a priori}
into account,
the MCTDH method has been specified
for systems of identical particles,
which opened up interesting possibilities to treat larger systems.
First MCTDHF -- the fermionic version of MCTDH --
was developed by three independent groups \cite{MCTDHF1,MCTDHF2,MCTDHF3}.
Shortly after,
MCTDHB -- the bosonic version of MCTDH --
was developed in \cite{MCTDHB0,MCTDHB1}.
For applications of MCTDHF 
to laser-matter interaction and other few-fermion problems see, e.g.,
Refs.~\cite{applF1,applF2,applF3,applF4,applF5,applF6,applF7,applF7m5,applF8,applF9},
where the last work combines optimal control theory with MCTDHF.
For applications of MCTDHB 
to Bose-Einstein condensates see, e.g.,
Refs.~\cite{applB1,applB2,applB3,applB4,applB5},
where the last two works combine optimal 
control theory with MCTDHB.

Since the seminal paper of L\"owdin \cite{Lowdin},
reduced density matrices and particularly reduced two-body density
matrices have been a lively field of research, see, e.g.,
Refs.~\cite{Slava,MAZZ1,MAZZ2,MAZZ3,MAZZ4,MAZZ5,MAZZ6}.
Reduced one-body density matrices 
are an inherent 
part of the MCTDH \cite{cpl,jcp,review,book}.
In the present context,
reduced one- and two-body density matrices
were first used to derive the
static self-consistent theory for bosons,
the multiconfigurational Hartree for bosons (MCHB) in \cite{MCHB}.
Thereafter,
MCTDHB and MCTDHF were formulated in a unified manner
by employing reduced one-, two- \cite{unified} and three-body \cite{book}
density matrices.
Further specification of MCTDH to mixtures of 
two kinds of identical particles
(MCTDH-FF for Fermi-Fermi mixtures;
MCTDH-BF for Bose-Fermi mixtures;
and
MCTDH-BB for Bose-Bose mixtures)
was put forward in \cite{MCTDHX}.
All the above developments made use of the
fact that the mean-field operators in
the traditional MCTDH can be factorized to
products of reduced density matrices 
times one-body operators. 
Finally, 
we mention that
MCTDH has been extended to systems with particle conversion
(termed MCTDH-{\it conversion}), 
where particles of one kind can 
convert to another kind \cite{conversion}.

A breakthrough in the formulation \cite{mapping,3well} and implementation \cite{package} 
of MCTDHB has stemmed from a general Combinadic-based mapping
of bosonic (and fermionic) operators in Fock space.
In this formulation,
the direct 
calculation of the matrix representation of
the Hamiltonian in the (huge) multiconfigurational 
space is abandoned,
and is replaced by the action of one-body and two-body 
density operators on the multiconfigurational wave-function.
The operation of the various density operators can be
performed in parallel \cite{package},
which further accelerates 
the performance of the algorithm.
This brings us closer to the topic and 
contents of the present work.

Two-body interaction is the most basic interaction in an interacting (quantum) system.
When the particles comprising the quantum system have internal structure,
higher-order interactions (forces) may come into play.
For instance, in nuclear physics it has long been accepted that three-body interactions
are necessary to fully understand the structure of nuclei, see, e.g. \cite{nuc3b,nuc3a}.
Much more recently, and in the context of another field,
the proposition to utilize cold polar molecules to engineer
(condensed-matter) systems with three-body interactions
has been made \cite{cold3}.
So, the motivation to study the non-equilibrium dynamics
of systems with up to three-body forces is clear.

But why study the quantum dynamics of a mixture of three kinds of identical particles?
Are such systems present in nature? 
In the cold-atom world,
the plurality of atoms is one of the most important ingredients
experimentalists (and theorists)
have at their disposal.
For instance,
the element Yb has
seven stable isotopes
(5 bosonic and 2 fermionic isotopes).
Yb has been envisaged to play an instrumental
role in realizing various interesting ultra-cold mixtures 
(see Ref.~\cite{Yb} for a realization of a 
Bose-Einstein condensate with $^{170}$Yb atoms
and the discussion therein).
More recently,
a quantum degenerate 
Fermi-Fermi mixture of $^6$Li-$^{40}$K atoms
coexisting with a Bose-Einstein Condensate
of $^{87}$Rb atoms were realized \cite{TH_2008}, 
as well as a triply quantum-degenerate mixture of 
bosonic $^{41}$K atoms 
and two fermionic $^{40}$K and $^6$Li atoms \cite{MZ_2011}.
Hence,
mixtures of three kinds of identical particles
have been created in the lab.

All the above dictate the purposes and contents of the present work.
The MCTDH method for
systems with three kinds of identical
particles interacting via all combinations of two- and three-body forces is derived,
and the resulting equations-of-motion are briefly discussed.
All four possible mixtures (Fermi-Fermi-Fermi,
Bose-Fermi-Fermi, Bose-Bose-Fermi and Bose-Bose-Bose) 
are presented in a unified manner.
Particular attention is paid to representing the 
coefficients' part of the equations-of-motion
in a compact recursive form 
in terms of one-body density operators only.
The recursion utilizes the recently proposed 
Combinadic-based mapping \cite{mapping} 
which has already been successfully applied 
and implemented within MCTDHB \cite{package}.
Our work sheds new light on the
representation of the coefficients'
part in MCTDHF and MCTDHB 
without resorting to the 
matrix elements of the many-body Hamiltonian 
with respect to the time-dependent configurations,
and suggests a recipe for 
efficient implementation of 
the theory derived here for mixtures
which is suitable for parallelization.

The structure of the paper is as follows.
In Sec.~\ref{SEC2} we present the building bricks
of the theory by reconstructing MCTDHF and MCTDHB.
In Sec.~\ref{SEC3} we assemble from
these ingredients the multiconfigurational
time-dependent Hartree method for mixtures
of three kinds 
of identical particles
interacting via up to three-body forces.
A brief summary and outlook are given in Sec.~\ref{SEC4}.
Finally,
we collect in Appendixes \ref{appendix_A}-\ref{appendix_C} 
for completeness and ease
of presentation of the main text 
various quantities appearing and needed
in the derivation.
The paper and the Appendixes are detailed and
intended also to serve as a guide for 
the implementation of the equations-of-motion.
The reconstruction of
MCTDHF and MCTDHB is given in sufficient detail.
This allows us to defer to the Appendixes
much of the lengthly formulas
used later on for the mixtures.

\section{Building bricks: Reconstructing MCTDHF and MCTDHB}\label{SEC2}

\subsection{From basic ingredients to mapping}\label{SEC2.1} 

Our starting point is the many-body Hamiltonian of
$N_A$ interacting identical particles of type $A$:
\beqn\label{ham}
 & & \hat H^{(A)} = \hat h^{(A)} + \hat W^{(A)} +  \hat U^{(A)} = 
 \int d\x \bigg\{ \hat{\mathbf \Psi}^\dag_A(\x) \hat h^{(A)}(\x) \hat{\mathbf \Psi}_A(\x) + \nonumber \\
 &+& \frac{1}{2} \int d\x' \bigg[ \hat{\mathbf \Psi}^\dag_A(\x) \hat{\mathbf \Psi}^\dag_A(\x') \hat W^{(A)}(\x,\x') 
 \hat{\mathbf \Psi}_A(\x') \hat{\mathbf \Psi}_A(\x) + \\
 &+& \frac{1}{3} \int d\x''  \hat{\mathbf \Psi}^\dag_A(\x) \hat{\mathbf \Psi}^\dag_A(\x') \hat{\mathbf \Psi}^\dag_A(\x'') 
 \hat W^{(A)}(\x,\x',\x'')  \hat{\mathbf \Psi}_A(\x'') \hat{\mathbf \Psi}_A(\x') 
\hat{\mathbf \Psi}_A(\x) \bigg] \bigg\}, \nonumber \
\eeqn
where $\hat h^{(A)}$ is the one-body part, 
$\hat W^{(A)}$ the two-body part
and $\hat U^{(A)}$ the three-body part.
The operators $\hat h^{(A)}$, $\hat W^{(A)}$ and $\hat U^{(A)}$
can generally be time-dependent.

We use the time-independent field operator expanded by time-dependent orbitals:
\beq\label{field}
 \hat{\mathbf \Psi}_A(\x) = \sum_k \hat a_k(t)\phi_k(\x,t),
\eeq
where the annihilation and creation operators obey
the usual fermionic/bosonic anti/commutation relations,
$\hat a_q(t) \hat a_k^\dag(t) \pm \hat a_k^\dag(t) \hat a_q(t) = \delta_{kq}$. 
Correspondingly,
the field operator obeys the anti/commutation relations,
$\hat{\mathbf \Psi}_A(\x) \left\{\hat{\mathbf \Psi}_A(\x')\right\}^\dag \pm
\left\{\hat{\mathbf \Psi}_A(\x')\right\}^\dag \hat{\mathbf \Psi}_A(\x) = \delta(\x-\x')$.
Here and hereafter the upper sign refers
to fermions and the lower to bosons.
The coordinate $\x \equiv \{\r, \sigma\}$
stands for spatial degrees of freedom and spin,
if present.
Thus, the shorthand notations
$\delta(\x-\x')=\delta(\r-\r')\delta_{\sigma,\sigma'}$
and $\int d\x \equiv \int d\r \sum_\sigma$
are implied throughout this work.
Furthermore,
we do not denote explicitly
the dependence of quantities on time when unambiguous.

Plugging the expansion (\ref{field}) into the many-body
Hamiltonian (\ref{ham}) one gets: 
\beq\label{ham2nd}
  \hat H^{(A)} =
 \sum_{k,q} h^{(A)}_{kq} \hat \rho^{(A)}_{kq}
 + \frac{1}{2} \sum_{k,s,q,l} W^{(A)}_{ksql} \hat \rho^{(A)}_{kslq} 
 + \frac{1}{6}\sum_{k,s,p,r,l,q} U^{(A)}_{kspqlr} \hat \rho^{(A)}_{ksprlq},
\eeq
where the matrix elements with respect to the orbitals $\left\{\phi_k(\x,t)\right\}$ are given by:
\beqn\label{matrix_elements}
 h^{(A)}_{kq} &=& \int \phi_k^\ast(\x,t) \hat h^{(A)}(\x) \phi_q(\x,t) d\x, \nonumber \\
W^{(A)}_{ksql} &=& \int \!\! \int \phi_k^\ast(\x,t) \phi_s^\ast(\x',t) \hat W^{(A)}(\x,\x')
 \phi_q(\x,t) \phi_l(\x',t) d\x d\x', \nonumber \\
U^{(A)}_{kspqlr} &=&  \int \!\! \int \!\! \int \phi_k^\ast(\x,t) \phi_s^\ast(\x',t)  
\phi_p^\ast(\x'',t)  \hat U^{(A)}(\x,\x',\x'') \times \nonumber \\
&\times& \phi_q(\x,t) \phi_l(\x',t) \phi_r(\x'',t) d\x d\x' d\x''. \
\eeqn
In (\ref{ham2nd}), 
we introduce the one-body density operators
\beq\label{density_oper_1B}
 \hat \rho^{(A)}_{kq} = \hat a_k^\dag \hat a_q,
\eeq
as well as the two- and three-body density operators
\beqn\label{density_oper_2B_3B}
& & \hat \rho^{(A)}_{kslq} = \hat a_k^\dag \hat a_s^\dag \hat a_l \hat a_q = 
\pm \hat \rho^{(A)}_{kq} \delta_{sl} \mp \hat \rho^{(A)}_{kl} \hat \rho^{(A)}_{sq}, \nonumber \\
& & \hat \rho^{(A)}_{ksprlq} =  \hat a_k^\dag \hat a_s^\dag \hat a_p^\dag \hat a_r  \hat a_l \hat a_q =
\pm \hat \rho^{(A)}_{kslq} \delta_{pr} - \hat \rho^{(A)}_{ksrq} 
\delta_{pl} + \hat \rho^{(A)}_{ksrl} \hat \rho^{(A)}_{pq}.
\eeqn
The reason for 
this choice of notation with density operators
in (\ref{ham2nd}) will 
become clear below.
We see that the two-body density operators $\left\{\hat \rho^{(A)}_{kslq}\right\}$
can be written as products
of the one-body density operators, 
and that the three-body density operators $\left\{\hat \rho^{(A)}_{ksprlq}\right\}$ 
can be written 
as products of the two- and one-body density operators,
and so on, recursively.
Hence, 
the one-body density
operators $\left\{\hat \rho^{(A)}_{kq}\right\}$ in (\ref{density_oper_1B}) 
are our basic building bricks.

The many-body wave-function is expanded by time-dependent
configurations (determinants $\left|\i;t\right>$ for fermions, permanents $\left|\n;t\right>$ for bosons)
assembled by distributing the $N_A$ particles over
the $M_A$ time-dependent orbitals introduced in 
the expansion (\ref{field}).
For fermions we write \cite{mapping}:
\beq\label{MCTDHF_ansatz}
\left|\Psi^{(A)}(t)\right> =
\sum_{\{\i\}} C_{\i}(t) \left|\i;t\right> \equiv 
\sum^{N^{(A)}_{\mathit{conf}}}_{J_A=1} C_{J_A}(t) \left|J_A;t\right>,
\eeq
where the address $J_A$ is defined as follows:
\beq\label{I_numbering}
J_A \equiv J_A(\i)= 1 + \sum_{j=1}^{M_A-N_A}\binom{M_A-i_j}{M_A-N_A+1-j},
\eeq
whereas for bosons we write \cite{mapping}:
\beq\label{MCTDHB_ansatz}
\left|\Psi^{(A)}(t)\right> =
\sum_{\{\n\}} C_{\n}(t) \left|\n;t\right> \equiv 
\sum^{N^{(A)}_{\mathit{conf}}}_{J_A=1} C_{J_A}(t) \left|J_A;t\right>, 
\eeq
where the address $J_A$ is defined as follows:
\beq\label{J_numbering}
 J_A \equiv J_A(\n) = 1 + \sum_{k=1}^{M_A-1} \binom{N_A+M_A-1-k-\sum_{l=1}^{k}n_l}{M_A-k}.
\eeq
The notation used in (\ref{MCTDHF_ansatz}-\ref{J_numbering}) 
follows the 
Combinadic-based addressing 
scheme of configurations introduced in \cite{mapping}.
For fermions we enumerate configurations by holes, $\i = (i_1,i_2,\ldots,i_j=q,\ldots,i_{M_A-N_A})$ and
$\i^{kq} = (i_1,i_2,\ldots,i_l=k,\ldots,i_{M_A-N_A})$,
whereas for bosons we enumerate configurations by particles,
$\n = (n_1,\ldots,n_k,\ldots,n_q,\ldots,n_{M_A})$ and
$\n^{kq} = (n_1,\ldots,n_k-1,\ldots,n_q+1,\ldots,n_{M_A})$.
The index $J_A$ is termed ``address" 
because it is an integer uniquely identifying a configuration which is described 
by the positions of the holes $\i$ (for fermions) or the occupation numbers $\n$ (for bosons).
For more details of the Combinadic-based mapping and particularly 
the connection between the bosonic occupation numbers
and the positions of the fermionic holes see \cite{mapping}.

For our requirements, 
we will need the result of the operation of
the basic building bricks
onto the state vector,
namely, the operation of the one-body density 
operators $\left\{\hat \rho^{(A)}_{kq}\right\}$
onto $\left|\Psi^{(A)}(t)\right>$.
Thus we have:
\beq\label{O_den}
 \hat \rho^{(A)}_{kq} \left|\Psi^{(A)}(t)\right> =  \hat \rho^{(A)}_{kq} \sum^{N^{(A)}_{\mathit{conf}}}_{J_A=1} 
C_{J_A}(t) \left|J_A;t\right> \equiv % =
  \sum^{N^{(A)}_{\mathit{conf}}}_{J_A=1} C^{\hat \rho^{(A)}_{kq}}_{J_A}(t) \left|J_A;t\right>.
\eeq
For fermions we have the following relations \cite{mapping}:
\beqn\label{basic_mapping_F}
\!\!\!\!\!\!\!\! C^{\hat \rho^{(A)}_{kq}}_{J_A}(t) \equiv C^{\hat \rho^{(A)}_{kq}}_{J_A(\i)}(t) &=& 
\left \{
\begin{matrix}
C_{J_A(\i^{kq})}(t) \times (-1)^{d(\i^{kq})}; & \ k \ne q, \ k \in \i^{kq}, \ q \not\in \i^{kq}\\
C_{J_A(\i)}(t); & \ k = q, \ k \not\in \i\\
0; & \ {\mathrm{otherwise}}  \\
\end{matrix}
\right., \ 
\eeqn
where the distance between the $i_j$-th hole of $\i$ at orbital $q$ and the $i_l$-th hole of $\i^{kq}$ 
at orbital $k$ is
given by $d(\i^{kq}) = |k-q| - |j-l| - 1$
[equivalently, $d(\i^{kq}) = \sum_{p \in (k,q)} n_p$ simply enumerates how many fermions are there between
the $k$-th and $q$-th orbitals].
For bosons we have the following relations \cite{mapping}:
\beqn\label{basic_mapping_B}
C^{\hat \rho^{(A)}_{kq}}_{J_A}(t) \equiv C^{\hat \rho^{(A)}_{kq}}_{J_A(\n)}(t) &=& 
\left \{
\begin{matrix}
C_{J_A(\n^{kq})}(t) \times \sqrt{n_k} \sqrt{n_q +1}; & \ k \ne q \\
C_{J_A(\n)}(t) \times n_k; & \ k = q \\
\end{matrix}
\right., \
\eeqn
which concludes our exposition of the Combinadic-based mapping 
and assembly of
the operations of the basic building bricks $\left\{\hat \rho^{(A)}_{kq}\right\}$
on the many-body wave-function $\left|\Psi^{(A)}(t)\right>$.
From Eqs.~(\ref{density_oper_1B},\ref{density_oper_2B_3B}) we see how
to use the one-body (basic) building bricks $\left\{\hat \rho^{(A)}_{kq}\right\}$
to assemble higher-body operators.
In particular we find:
\beqn\label{basic_mapping_2B_3B}
 & & C^{\hat \rho^{(A)}_{kslq}}_{J_A}(t) = \pm \delta_{sl} C^{\hat \rho^{(A)}_{kq}}_{J_A}(t) 
\mp {C^{\hat \rho^{(A)}_{sq}}_{J_A}}^{\hat \rho^{(A)}_{kl}}\!\!(t), \nonumber \\
 & & C^{\hat \rho^{(A)}_{ksprlq}}_{J_A}(t) = \pm \delta_{pr} C^{\hat \rho^{(A)}_{kslq}}_{J_A}(t)
 - \delta_{pl} C^{\hat \rho^{(A)}_{ksrq}}_{J_A}(t) + 
{C^{\hat \rho^{(A)}_{pq}}_{J_A}}^{\hat \rho^{(A)}_{ksrl}}\!(t). \
\eeqn
The meaning of the two levels of density operators in the 
superscripts of the coefficients 
$C^{\hat \rho^{(A)}_{kslq}}_{J_A}(t)$ 
and 
$C^{\hat \rho^{(A)}_{ksprlq}}_{J_A}(t)$,
resulting from higher-body operators 
in (\ref{basic_mapping_2B_3B}),
is that the lower-level density operator is multiplied on the many-body wave-function first,
and the upper-level 
density operator is multiplied thereafter on the result.

The key ingredient in the utilization of the Lagrangian formulation \cite{MCTDHB1,LF1,LF2} 
of the (Dirac-Frenkel \cite{DF1,DF2}) 
time-dependent variational principle to derive the equations-of-motion is
the evaluation of matrix elements with respect to the
multiconfigurational wave-function $\left|\Psi^{(A)}(t)\right>$.
This will be utilized in the next subsection \ref{SEC2.2}.
For the moment,
we would like to prescribe how
such matrix elements with respect to $\left|\Psi^{(A)}(t)\right>$
are to be evaluated.

Consider the operator $\hat O^{(A)}$,
which can be a one-body operator, two-body operator, three-body operator, etc.
Then, we express and compute the expectation value of $\hat O^{(A)}$ 
with respect to $\left|\Psi^{(A)}(t)\right>$
as follows \cite{mapping}:
\beq\label{expectation}
\left<\Psi^{(A)}(t)\left| \hat O^{(A)} \right|\Psi^{(A)}(t)\right> =  
\left<\Psi^{(A)}(t)\left| \left\{ \hat O^{(A)} \right|\Psi^{(A)}(t)\right> \right\} = 
\sum^{N^{(A)}_{\mathit{conf}}}_{J_A=1} C^\ast_{J_A}(t) C^{\hat O^{(A)}}_{J_A}(t), 
\eeq
where
\beq\label{O_Psi}
 \hat O^{(A)} \left|\Psi^{(A)}(t)\right> = 
\hat O^{(A)} \sum^{N^{(A)}_{\mathit{conf}}}_{J_A=1} C_{J_A}(t) \left|J_A;t\right> \equiv 
  \sum^{N^{(A)}_{\mathit{conf}}}_{J_A=1} C^{\hat O^{(A)}}_{J_A}(t) \left|J_A;t\right>.
\eeq
In particular,
for a one-body operator, 
 $\hat O^{(A)} = \sum_{k,q} O^{(A)}_{kq} \hat \rho^{(A)}_{kq}$, we get:
\beq\label{C_one}
 C^{\hat O^{(A)}}_{J_A}(t) = \sum_{k,q}^{M_A} O^{(A)}_{kq}  C^{\hat \rho^{(A)}_{kq}}_{J_A}(t),
\eeq
for a two-body operator, $\hat O^{(A)} = \frac{1}{2} \sum_{k,s,q,l} O^{(A)}_{ksql} \hat \rho^{(A)}_{kslq}$, 
we get from (\ref{basic_mapping_2B_3B}):
\beqn\label{C_two}
 C^{\hat O^{(A)}}_{J_A}(t) &=& \frac{1}{2} 
\sum_{k,s,q,l}^{M_A} O^{(A)}_{ksql} C^{\hat \rho^{(A)}_{kslq}}_{J_A}(t) = \nonumber \\ 
 &=& \frac{1}{2} \sum_{k,s,q,l}^{M_A} O^{(A)}_{ksql} \left[ \pm \delta_{sl} C^{\hat \rho^{(A)}_{kq}}_{J_A}(t) 
\mp {C^{\hat \rho^{(A)}_{sq}}_{J_A}}^{\hat \rho^{(A)}_{kl}}\!\!(t) \right], \ 
\eeqn
and for a 
three-body operator, $\hat O^{(A)} = \frac{1}{6} 
\sum_{k,s,p,r,l,q} O^{(A)}_{kspqlr} \hat \rho^{(A)}_{ksprlq}$, 
we get from (\ref{basic_mapping_2B_3B}):
\beqn\label{C_three}
 & & 
C^{\hat O^{(A)}}_{J_A}(t) = \frac{1}{6} \sum_{k,s,p,r,l,q}^{M_A}
 O^{(A)}_{kspqlr} C^{\hat \rho^{(A)}_{ksprlq}}_{J_A}(t) = \\
  &=& \frac{1}{6} \sum_{k,s,p,r,l,q}^{M_A} O^{(A)}_{kspqlr} \left[ \pm \delta_{pr} C^{\hat \rho^{(A)}_{kslq}}_{J_A}(t)
 - \delta_{pl} C^{\hat \rho^{(A)}_{ksrq}}_{J_A}(t) + 
{C^{\hat \rho^{(A)}_{pq}}_{J_A}}^{\hat \rho^{(A)}_{ksrl}}\!(t)
 \right]. \nonumber \
\eeqn
Finally and generally, 
the result of a sum of (operations of) operators, e.g., $\hat O_1^{(A)} + \hat O_2^{(A)}$,
on $\left|\Psi^{(A)}(t)\right>$
translates to the sum of the respective coefficients \cite{mapping}:
\beq\label{operators_sum}
 C^{\hat O_1^{(A)} + \hat O_2^{(A)}}_{J_A}(t) = C^{\hat O_1^{(A)}}_{J_A}(t) + C^{\hat O_2^{(A)}}_{J_A}(t).
\eeq 
These compact relations resting on one-body density operators only
[the two-body density operators in (\ref{C_three})
are assembled from one-body density 
operators according to (\ref{density_oper_1B},\ref{density_oper_2B_3B})] 
will be used to reformulate MCTDHF and MCTDHB
in a recursive manner in the following subsection \ref{SEC2.2}.

\subsection{Equations-of-motion utilizing one-body density operators 
and Combinadic-based mapping}\label{SEC2.2} 

We can derive (reconstruct)
the MCTDHF and MCTDHB equations-of-motion,
taking into account {\it a-priori}
that matrix elements of the form of (\ref{expectation}) enter the variational formulation.
Within the Lagrangian formulation \cite{MCTDHB1,LF1,LF2} of the (Dirac-Frenkel \cite{DF1,DF2}) 
time-dependent variational principle,
the action functional of the time-dependent
many-particle Schr\"odinger equation takes
on the following form:
\beqn\label{func_basic}
 & &  S\left[\left\{C_{J_A}(t)\right\},\left\{\phi_k(\x,t)\right\}\right] =
\int dt \Bigg\{\left< \Psi^{(A)}(t) \left| \hat H^{(A)} - i\frac{\partial}{\partial t}\right| \Psi^{(A)}(t)\right>
 - \nonumber \\
 & & \qquad - \sum_{k,j}^{M_A} \mu_{kj}^{(A)}(t) \left[\left<\phi_k \left|\right.\phi_j\right> - \delta_{kj}\right]
- \varepsilon^{(A)}(t) \left[\sum_{J_A=1}^{N^{(A)}_{\mathit{conf}}} \left|C_{J_A}(t)\right|^2 - 1 \right]\Bigg\}, \
\eeqn
where the time-dependent Lagrange multipliers
$\left\{\mu_{kj}^{(A)}(t)\right\}$ are introduced
to guarantee the orthonormality
of the orbitals at all times.
Furthermore,
they enable one to first evaluate the expectation
value of $\hat H^{(A)} - i\frac{\partial}{\partial t}$
with respect to $\left|\Psi^{(A)}(t)\right>$
and then to perform the variation,
which is precisely what is needed in
order to exploit the Combinadic-based 
mapping \cite{mapping} {\it a-priori}
in the derivation of the equations-of-motion.
The (here redundant) time-dependent Lagrange multiplier $\varepsilon^{(A)}(t)$
ensures normalization of the expansion coefficients at all times,
and would resurface in the static theory 
in the case of the 
imaginary-time-propagation formulation.

To perform the variation of the action functional with 
respect to the coefficients, 
we express the expectation value
$\left< \Psi^{(A)}(t) \left| \hat H^{(A)} - i\frac{\partial}{\partial t}\right| \Psi^{(A)}(t)\right>$
following the Combinadic-based 
mapping \cite{mapping} 
and the compact expression in Eq.~(\ref{expectation}):
\beq\label{expectation_H_C}
\left<\Psi^{(A)}(t)\left| \hat H^{(A)} - i\frac{\partial}{\partial t} \right|\Psi^{(A)}(t)\right> =  
\sum^{N^{(A)}_{\mathit{conf}}}_{J_A=1}
 C^\ast_{J_A}(t) \left[ C^{\hat H^{(A)} - i\frac{\partial}{\partial t}^{(A)}}_{J_A}\!(t) - i \dot C_{J_A}(t) \right]. 
\eeq
Representation (\ref{expectation_H_C}) makes
it clear what the variation with respect to the
coefficients $\left\{C^\ast_{J_A}(t)\right\}$ would lead to.
When this 
variation 
is performed explicitly,
one immediately finds:
\beq\label{C_gen}
 C^{\hat H^{(A)} - i\frac{\partial}{\partial t}^{(A)}}_{J_A}\!(t) = i \dot C_{J_A}(t), \qquad \forall J_A.
\eeq
The meaning of $i\frac{\partial}{\partial t}^{(A)}$
is that the time-derivative is a one-body operator in the
$A$-species Fock (and orbital) space.
According to the rules of the previous subsection \ref{SEC2.1},
the left-hand-side of Eq.~(\ref{C_gen}) is given by the sum of its
one-, two- and three-body constituents:
\beq\label{SE_C_gen}
 C^{\hat H^{(A)} - i\frac{\partial}{\partial t}^{(A)}}_{J_A}\!(t) =
 C^{\hat h^{(A)} - i\frac{\partial}{\partial t}^{(A)}}_{J_A}\!(t) +
 C^{\hat W^{(A)}}_{J_A}(t) + C^{\hat U^{(A)}}_{J_A}(t).
\eeq

The invariance of $\left|\Psi^{(A)}(t)\right>$
to unitary transformations of the orbitals,
compensated by the `reverse' transformations
of the orbitals is well-known \cite{cpl,jcp,MCTDHB1} and
can be represented as follows:
$\left|\Psi^{(A)}(t)\right> = \sum^{N^{(A)}_{\mathit{conf}}}_{J_A=1} C_{J_A}(t) \left|J_A;t\right> =$\break\hfill
$\sum^{N^{(A)}_{\mathit{conf}}}_{J_A=1} \overline{C}_{J_A}(t) \overline{\left|J_A;t\right>}$,
with obvious notation.
This invariance can
be utilized to bring the equations-of-motion into a simpler form
(see, in particular, the discussion below on the orbitals' part).
Primarily, the differential conditions first introduced
by the MCTDH founders \cite{cpl,jcp}:
\beq\label{diff_con_A}
 \left\{i\frac{\partial}{\partial t}^{(A)}\right\}_{kq} \equiv
 i\left<\phi_k \left|\dot\phi_q\right>\right. = 0, \ \ k,q=1,\ldots,M_A,
\eeq
come 
out explicitly from such a unitary transformation \cite{MCTDHB1,conversion} and
straightforwardly 
lead 
in the case of the
equations-of-motion for the coefficients to:
\beqn\label{C_gen_phi_phidot}
& & C^{\hat H^{(A)}}_{J_A}(t) = i \dot C_{J_A}(t), \qquad \forall J_A, \nonumber \\
& & C^{\hat H^{(A)}}_{J_A}(t) =
 C^{\hat h^{(A)}}_{J_A}(t) + C^{\hat W^{(A)}}_{J_A}(t) + C^{\hat U^{(A)}}_{J_A}(t). \
\eeqn
For the general form of the differential conditions, 
Eq.~(\ref{diff_con_A}), 
see the literature \cite{review,book}.
We remark that a particular interesting representation 
(put forward and utilized so far for distinguishable degrees-of-freedom only) 
of the differential conditions 
can be made in order 
to propagate the systems' 
natural orbitals \cite{Uwe_nat1,Uwe_nat2}.

In MCTDHF and MCTDHB
the integration of the coefficients' part in time
is performed (for unitary time-evolution) 
by the short iterative Lanczos (SIL) algorithm \cite{SIL}.
We remark on the numerical implementation of Eq.~(\ref{C_gen_phi_phidot})
within SIL propagation \cite{package}. 
For the SIL one needs to operate with powers of $\hat H$ onto the many-particle wave-function
and construct the $K$-dimensional Krylov subspace:
$\left\{\left|\Psi^{(A)}(t)\right>, \hat H^{(A)}\left|\Psi^{(A)}(t)\right>,\ldots,
\hat{H}^{(A)}\strut^{K-1} \left|\Psi^{(A)}(t)\right> \right\}$.
In the language of the Combinadic-based mapping of coefficients 
and utilizing the recipe of 
how to operate with operators on the many-particle wave-function 
discussed above \cite{mapping}, 
this construction translates to:
$\left\{C_{J_A}(t), C^{\hat H^{(A)}}_{J_A}(t),{C^{\hat H^{(A)}}_{J_A}}^{\hat H^{(A)}}\!(t),\ldots\right\}$.

Let us now move to the equations-of-motion for the orbitals
$\left\{\phi_k(\x,t)\right\}$.
For this,
the expectation value of the many-body Hamiltonian $\hat H^{(A)}$
with respect to $\left|\Psi^{(A)}(t)\right>$ has to be expressed
in a form which allows for variation with respect to the orbitals,
namely as an 
explicit function of the quantities (integrals) $h^{(A)}_{kq}$,
$W^{(A)}_{ksql}$ and $U^{(A)}_{kspqlr}$ in (\ref{matrix_elements}).
The result reads:
\beqn\label{expectation_H3_phi}
\!\!\!\!\!\!\!\! & & 
\left<\Psi\left|\hat H^{(A)} - i\frac{\partial}{\partial t} \right|\Psi\right> =
 \sum_{k,q=1}^{M_A} \rho^{(A)}_{kq} \left[ h^{(A)}_{kq} - 
\left\{i\frac{\partial}{\partial t}^{(A)}\right\}_{kq} \right] + \\
 & & 
 + \frac{1}{2}\sum_{k,s,l,q=1}^{M_A} \rho^{(A)}_{kslq} W^{(A)}_{ksql}
 + \frac{1}{6}\sum_{k,s,p,r,l,q=1}^{M_A} \rho^{(A)}_{ksprlq} U^{(A)}_{kspqlr}
- \sum^{N^{(A)}_{\mathit{conf}}}_{J_A=1}
 i C^\ast_{J_A}(t) \dot C_{J_A}(t). \nonumber \
\eeqn
The expectation values of the density operators $\hat \rho^{(A)}_{kq}$,
$\hat \rho^{(A)}_{kslq}$ and $\hat \rho^{(A)}_{ksprlq}$
with respect to $\left|\Psi^{(A)}(t)\right>$
(resulting from the expectation value of the Hamiltonian 
with respect to many-particle wave-function) are computed 
following Eq.~(\ref{expectation}):
\beqn\label{denisty_matrx_element}
& & \rho^{(A)}_{kq} =  \sum^{N^{(A)}_{\mathit{conf}}}_{J_A=1} C^\ast_{J_A}(t) C^{\hat \rho^{(A)}_{kq}}_{J_A}(t), 
\qquad  \rho^{(A)}_{kslq} = \sum^{N^{(A)}_{\mathit{conf}}}_{J_A=1} C^\ast_{J_A}(t) C^{\hat \rho^{(A)}_{kslq}}_{J_A}(t), 
\nonumber \\
& & \qquad \qquad 
 \rho^{(A)}_{ksprlq} = \sum^{N^{(A)}_{\mathit{conf}}}_{J_A=1} C^\ast_{J_A}(t) C^{\hat \rho^{(A)}_{ksprlq}}_{J_A}(t), \
\eeqn
where the 
coefficients 
$C^{\hat \rho^{(A)}_{kq}}_{J_A}(t)$, $C^{\hat \rho^{(A)}_{kslq}}_{J_A}(t)$
and $C^{\hat \rho^{(A)}_{ksprlq}}_{J_A}(t)$
are given in Eqs.~(\ref{basic_mapping_F},\ref{basic_mapping_B}) 
and (\ref{basic_mapping_2B_3B}), 
respectively.
We collect the expectation values
of the one-body density operators
as the matrix $\brho^{(A)}(t)=\left\{\rho^{(A)}_{kq}(t)\right\}$.

One should remember that
the expectation values 
of two- and three-body
density operators can generally not be factorized
into products of expectation values of one-body density operators.
For instance (and in the language of the Combinadic-based mapping
of coefficients),
$C^{\hat \rho^{(A)}_{kslq}}_{J_A}(t) = \pm \delta_{sl} C^{\hat \rho^{(A)}_{kq}}_{J_A}(t) 
\mp {C^{\hat \rho^{(A)}_{sq}}_{J_A}}^{\hat \rho^{(A)}_{kl}}\!\!(t)
\ne
 \pm \delta_{sl} C^{\hat \rho^{(A)}_{kq}}_{J_A}(t) 
 \mp C^{\hat \rho^{(A)}_{kl}}_{J_A}(t) C^{\hat \rho^{(A)}_{sq}}_{J_A}(t)$.
This is unlike the operation of the density operators themselves 
on the many-particle wave-function utilized above.

We can now perform 
the variation of 
$S\left[\left\{C_{J_A}(t)\right\},\left\{\phi_k(\x,t)\right\}\right]$
with respect to the orbitals.
This variation has been detailed
in the literature, see \cite{MCTDHB1,unified}, and we give here
the main steps in the derivation of the equations-of-motion
as far as they are needed for our needs later on.
Making use of the orthonormality relation between the 
time-dependent orbitals $\left\{\phi_k(\x,t)\right\}$,
we can solve for the Lagrange multipliers, 
$k,j = 1,\ldots,M_A$:
\beqn\label{MCTDHX_H3_mu}
& & \!\!\!\!\!\!\!\! \mu_{kj}^{(A)}(t) =  \\ 
& & \!\!\!\!\!\!\!\!  =
\left<\phi_j\left| 
\sum^{M_A}_{q=1} \left( \rho^{(A)}_{kq} \left[ \hat h^{(A)} 
 - i\frac{\partial}{\partial t}^{(A)} \right] + \sum^{M_A}_{s,l=1}\rho^{(A)}_{kslq} \hat W^{(A)}_{sl} 
 + \frac{1}{2}\sum_{s,p,r,l=1}^{M_A} \rho^{(A)}_{ksprlq} \hat U^{(A)}_{splr} \right) \right|\phi_q\right>. \nonumber \
\eeqn
The Lagrange multipliers $\left\{\mu_{kj}^{(A)}(t)\right\}$ can
be eliminated from the equations-of-motion which 
is achieved by the introduction 
of the projection operator:
\beq\label{project_A}
\hat {\mathbf P}^{(A)} = 1 - \sum_{u=1}^{M_A} \left|\phi_{u}\right>\left<\phi_{u}\right|.
\eeq
When this is done, 
we find the following equations-of-motion for the orbitals $\left\{\phi_j(\x,t)\right\}$,
$j=1,\ldots,M_A$:
\beqn\label{MCTDHX_P_P_H3_eom}
& &  \hat {\mathbf P}^{(A)} i\left|\dot\phi_j\right> =  \hat {\mathbf P}^{(A)} 
\Bigg[\hat h^{(A)} \left|\phi_j\right>  + \\
& & + \sum^{M_A}_{k,q=1}
  \left\{\brho^{(A)}(t)\right\}^{-1}_{jk}
\sum^{M_A}_{s,l=1}
\left(\rho^{(A)}_{kslq} \hat{W}^{(A)}_{sl}
+\frac{1}{2}\sum_{p,r=1}^{M_A} \rho^{(A)}_{ksprlq} \hat U^{(A)}_{splr} \right)
\left|\phi_q\right> \Bigg], \nonumber \
\eeqn
where
\beqn\label{TD_1B_2_3_potentials}
& & \hat W^{(A)}_{sl}(\x,t)=\int\phi_s^\ast(\x',t) \hat W^{(A)}(\x,\x') \phi_l(\x',t) d\x', \\
& & \hat U^{(A)}_{splr}(\x,t) = \int \!\! \int \phi_s^\ast(\x',t)
\phi_p^\ast(\x'',t)  \hat U^{(A)}(\x,\x',\x'') \phi_l(\x',t) \phi_r(\x'',t) d\x' d\x'', \nonumber
\eeqn
are local (for spin-independent interactions), 
time-dependent one-body potentials,
and $\dot \phi_j \equiv \frac{\partial\phi_j}{\partial t}$.

Utilizing the differential conditions (\ref{diff_con_A}) 
we can eliminate the projection operator  $\hat {\mathbf P}^{(A)}$ 
appearing on the left-hand-side of
Eq.~(\ref{MCTDHX_P_P_H3_eom}) and arrive at
the final result for the equations-of-motion of
the orbitals in MCTDHF and MCTDHB (see \cite{book,unified}), 
$j=1,\ldots,M_A$:
\beqn\label{MCTDHX_P_H3_eom}
& &  i\left|\dot\phi_j\right> =  \hat {\mathbf P}^{(A)} 
\Bigg[\hat h^{(A)} \left|\phi_j\right>  + \\
& & + \sum^{M_A}_{k,q=1}
  \left\{\brho^{(A)}(t)\right\}^{-1}_{jk}
\sum^{M_A}_{s,l=1}
\left(\rho^{(A)}_{kslq} \hat{W}^{(A)}_{sl}
+\frac{1}{2}\sum_{p,r=1}^{M_A} \rho^{(A)}_{ksprlq} \hat U^{(A)}_{splr} \right)
\left|\phi_q\right> \Bigg]. \nonumber \
\eeqn
Summarizing,
the coupled sets of equations-of-motion (\ref{C_gen_phi_phidot}) 
for the expansion coefficients and (\ref{MCTDHX_P_H3_eom}) 
for the orbitals 
constitute
the MCTDHF and MCTDHB methods,
where the one-body density operators 
(\ref{density_oper_1B},\ref{density_oper_2B_3B})
are employed as the basic building bricks 
in their construction and implementation.

We can also 
propagate the MCTDHF and MCTDHB 
equations-of-motion (\ref{C_gen_phi_phidot},\ref{MCTDHX_P_H3_eom})
in imaginary time and
arrive for time-independent Hamiltonians
at the corresponding self-consistent 
static theories, MCHF \cite{gen_MCHF1,gen_MCHF2} and MCHB \cite{MCHB}.
Thus,
setting $t \to -it$ into the coupled sets (\ref{C_gen},\ref{MCTDHX_P_P_H3_eom})
or into (\ref{C_gen_phi_phidot},\ref{MCTDHX_P_H3_eom}),
and translating back from the projection operator $\hat {\mathbf P}^{(A)}$
to the Lagrange multipliers $\left\{\mu_{kj}^{(A)}\right\}$,
the final result reads, $k=1,\ldots,M_A$:
\beqn\label{MCTDH_H3_stationary}
 & & \!\!\!\!\!\!\!\! 
\sum_{q=1}^{M_A} \left[ \rho^{(A)}_{kq} \hat h^{(A)} +
  \sum^{M_A}_{s,l=1} \left(\rho^{(A)}_{kslq} \hat{W}^{(A)}_{sl}
 + \frac{1}{2}\sum_{p,r=1}^{M_A} \rho^{(A)}_{ksprlq} \hat U^{(A)}_{splr}\right)
\right] \left|\phi_q\right> =
 \sum_{j=1}^{M_A} \mu_{kj}^{(A)} \left|\phi_j\right>, \nonumber \\
 & & 
\qquad \qquad 
C^{\hat H^{(A)}}_{J_A} = \varepsilon^{(A)} C_{J_A}, \qquad \forall J_A, \
\eeqn
where, making use of the normalization of the many-particle wave-function, 
$\varepsilon^{(A)}= \sum^{N^{(A)}_{\mathit{conf}}}_{J_A=1} C^\ast_{J_A} C^{\hat H^{(A)}}_{J_A}$
is the eigen-energy of the system.
Making use of the fact that the matrix of Lagrange multipliers
$\{\mu_{kj}^{(A)}\}$ is Hermitian (for stationary states)
and of 
the invariance property of the multiconfigurational wave-function
(to unitary transformations 
of the orbitals 
compensated by the `reverse' 
transformations
of the coefficients), 
one can transform Eq.~(\ref{MCTDH_H3_stationary})
to a representation where $\{\mu_{kj}^{(A)}\}$
is a diagonal matrix.

All in all,
we have formulated in the present section
the MCTDHF and MCTDHB equations-of-motion,
as well as their static 
variants MCHF and MCHB,
by (i) utilizing in a recursive manner one-body density operators only, 
and by (ii) employing {\it a priori} the Combinadic-based mapping 
formulation of Ref.~\cite{mapping} to evaluate matrix elements.
This sets up the tools to
put forward the MCTDH theory
for mixtures of three kinds of identical particles in the following Sec.~\ref{SEC3},
and to briefly 
discuss its structure and properties,
and how to implement it.

\section{Three kinds of identical particles: MCTDH-FFF, MCTDH-BFF, MCTDH-BBF and MCTDH-BBB}\label{SEC3}

In the present section we specify the MCTDH theory for mixtures of three 
kinds of identical particles, interacting with up to three-body forces. 
Before we get into the details of derivation and flood of equations, 
we would like to lay out a general scheme or flowchart that one can follow to handle 
similar or even more complex mixtures. 
Specifically, we need to assign a different set of time-dependent orthonormal orbitals to each and every species in the mixture. 
These orbitals are then used to assemble the time-dependent configurations 
(with determinants' parts for fermions and permanents' parts for bosons). 
The many-particle wave-function is thereafter assembled as a linear combination of 
all time-dependent configurations with time-dependent expansion coefficients. 
The many-particle Hamiltonian contains different terms: 
It contains intra-species terms and inter-species terms which consist of two-body, 
three-body and so on interactions. 
The main point in the representation of the Hamiltonian is the utilization of one-body density operators. 
In turn, all intra-species and inter-species interactions can be represented 
utilizing (products of) one-body density operators only.

The key step in the derivation of the equations-of-motion is the utilization of the 
Lagrangian formulation \cite{MCTDHB1,LF1,LF2} of the (Dirac-Frenkel \cite{DF1,DF2})
time-dependent variational principle with Lagrange multipliers for each species' orbitals, ensuring thereby 
the orthonormality of the orbitals for all times. In such a way, 
matrix-elements appear within the formulation explicitly, 
before the variation with respect to either the expansion coefficients 
or the orbitals is performed. The equations-of-motion for the expansion 
coefficients of the multiconfigurational wave-function are obtained by taking the 
variation of the action functional when it is expressed explicitly in terms of the expansion coefficients. 
The Combinadic-based mapping \cite{mapping} lifts the necessity to work with the huge matrix 
representation of the Hamiltonian with respect to the configurations, 
and allows one to efficiently perform operations on the vector of expansion coefficients directly. 
The equations-of-motion for the orbitals are obtained by taking the variation of the action 
functional when it is expressed explicitly in terms of the (integrals of the) orbitals. 
When this is performed, expectation values of the various density 
operators in the Hamiltonian (with respect to the many-particle wave-function) emerge 
which can be efficiently computed utilizing the Combinadic-based mapping \cite{mapping}.

\subsection{Additional ingredients 
for mixtures}\label{SEC3.1}

For a mixture of three kinds of identical particles,
$N_A$ particles of type $A$,
$N_B$ particles of type $B$
and
$N_C$ particles of type $C$,
we need now two additional field operators expanded by different complete
sets of time-dependent orbitals:
\beq\label{field_3Mix}
  \hat{\mathbf \Psi}_B(\y) = \sum_{k'} \hat b_{k'}(t) \psi_{k'}(\y,t), \qquad
 \hat{\mathbf \Psi}_C(\z) = \sum_{k''} \hat c_{k''}(t)\chi_{k''}(\z,t),
\eeq
where the field operator for the $A$-species particles $\hat{\mathbf \Psi}_A(\x)$
was first introduced 
and expanded in (\ref{field}). 
Note that each species can have
a different spin,
hence the explicit three distinct coordinates 
$\x$, $\y$ and $\z$.  
Field operators of distinct particles (can be chosen to) commute.

Our starting point is the many-body Hamiltonian 
of the most general 3-species mixture with up to 3-body 
interactions:
\beqn\label{ham_3mix_general}
& & \hat H^{(ABC)} =  \hat H^{(A)} + \hat H^{(B)} + \hat H^{(C)} + \hat W^{(AB)} + \hat W^{(AC)} + \hat W^{(BC)} + \\
& & + \hat U^{(AAB)} + \hat U^{(ABB)} + \hat U^{(AAC)} + \hat U^{(ACC)}
    + \hat U^{(BBC)} + \hat U^{(BCC)} + \hat U^{(ABC)}.  \nonumber \ 
\eeqn
Here, $\hat H^{(A)}$, $\hat H^{(B)}$ and $\hat H^{(C)}$ 
are the single-species Hamiltonians that can be read of (\ref{ham}).
The inter-species two-body interaction parts are given by:
\beqn\label{2_body_forces}
& &  \hat W^{(AB)} =  \int \!\! \int d\x  d\y \hat{\mathbf \Psi}^\dag_A(\x) \hat{\mathbf \Psi}^\dag_B(\y) 
  \hat W^{(AB)}(\x,\y) \hat{\mathbf \Psi}_B(\y) \hat{\mathbf \Psi}_A(\x), \nonumber \\
& & \hat W^{(AC)} =  \int \!\! \int d\x d\z \hat{\mathbf \Psi}^\dag_A(\x) \hat{\mathbf \Psi}^\dag_C(\z) 
  \hat W^{(AC)}(\x,\z) \hat{\mathbf \Psi}_C(\z) \hat{\mathbf \Psi}_A(\x), \nonumber \\
& & \hat W^{(BC)} =  \int \!\! \int d\y d\z \hat{\mathbf \Psi}^\dag_B(\y) \hat{\mathbf \Psi}^\dag_C(\z) 
  \hat W^{(BC)}(\y,\z) \hat{\mathbf \Psi}_C(\z) \hat{\mathbf \Psi}_B(\y). \
\eeqn
The 
inter-species three-body interaction parts,
resulting from the force between two identical particles and a 
third distinct particle, 
are given by:
\beqn\label{binary_3_body_forces}
\hat U^{(AAB)} &=&  \frac{1}{2} \int \!\! \int \!\! \int \!\! d\x d\x' d\y \hat{\mathbf \Psi}^\dag_A(\x) 
\hat{\mathbf \Psi}^\dag_A(\x') \hat{\mathbf \Psi}^\dag_B(\y) \hat U^{(AAB)}(\x,\x',\y) \times \nonumber \\
& & \times \hat{\mathbf \Psi}_B(\y) \hat{\mathbf \Psi}_A(\x') \hat{\mathbf \Psi}_A(\x), \nonumber \\
\hat U^{(ABB)} &=&  \frac{1}{2} \int \!\! \int \!\! \int \!\! d\x d\y d\y' \hat{\mathbf \Psi}^\dag_A(\x) 
\hat{\mathbf \Psi}^\dag_B(\y) \hat{\mathbf \Psi}^\dag_B(\y') \hat U^{(ABB)}(\x,\y,\y') \times \nonumber \\
& & \times \hat{\mathbf \Psi}_B(\y') \hat{\mathbf \Psi}_B(\y) \hat{\mathbf \Psi}_A(\x), \nonumber \\
\hat U^{(AAC)} &=&  \frac{1}{2} \int \!\! \int \!\! \int \!\! d\x d\x' d\z \hat{\mathbf \Psi}^\dag_A(\x) 
\hat{\mathbf \Psi}^\dag_A(\x') \hat{\mathbf \Psi}^\dag_C(\z) \hat U^{(AAC)}(\x,\x',\z) \times \nonumber \\
& & \times \hat{\mathbf \Psi}_C(\z) \hat{\mathbf \Psi}_A(\x') \hat{\mathbf \Psi}_A(\x), \nonumber \\
\hat U^{(ACC)} &=&  \frac{1}{2} \int \!\! \int \!\! \int \!\! d\x d\z d\z' \hat{\mathbf \Psi}^\dag_A(\x) 
\hat{\mathbf \Psi}^\dag_C(\z) \hat{\mathbf \Psi}^\dag_C(\z') \hat U^{(ACC)}(\x,\z,\z') \times \nonumber \\
& & \times \hat{\mathbf \Psi}_C(\z') \hat{\mathbf \Psi}_C(\z) \hat{\mathbf \Psi}_A(\x), \nonumber \\
\hat U^{(BBC)} &=&  \frac{1}{2} \int \!\! \int \!\! \int \!\! d\y d\y' d\z \hat{\mathbf \Psi}^\dag_B(\y) 
\hat{\mathbf \Psi}^\dag_B(\y') \hat{\mathbf \Psi}^\dag_C(\z) \hat U^{(BBC)}(\y,\y',\z) \times \nonumber \\
& & \times \hat{\mathbf \Psi}_C(\z) \hat{\mathbf \Psi}_B(\y') \hat{\mathbf \Psi}_B(\y), \nonumber \\
\hat U^{(BCC)} &=&  \frac{1}{2} \int \!\! \int \!\! \int \!\! d\y d\z d\z' \hat{\mathbf \Psi}^\dag_B(\y) 
\hat{\mathbf \Psi}^\dag_C(\z) \hat{\mathbf \Psi}^\dag_C(\z') \hat U^{(BCC)}(\y,\z,\z') \times \nonumber \\
& & \times \hat{\mathbf \Psi}_C(\z') \hat{\mathbf \Psi}_C(\z) \hat{\mathbf \Psi}_B(\y).
\eeqn
Finally, the inter-species three-body interaction part,
resulting from the force between three different particles 
 is given by:
\beqn\label{3_body_forces}
\hat U^{(ABC)} &=& \int \!\! \int \!\! \int \!\! d\x d\y d\z \hat{\mathbf \Psi}^\dag_A(\x) 
\hat{\mathbf \Psi}^\dag_B(\y) \hat{\mathbf \Psi}^\dag_C(\z) \hat U^{(ABC)}(\x,\y,\z) \times \nonumber \\
& & \times \hat{\mathbf \Psi}_C(\z) \hat{\mathbf \Psi}_B(\y) \hat{\mathbf \Psi}_A(\x).
\eeqn

When all the above are combined,
i.e., the field operators 
$\hat{\mathbf \Psi}_A(\x)$,
$\hat{\mathbf \Psi}_B(\y)$
and
$\hat{\mathbf \Psi}_B(\z)$
substituted into the various interaction terms,
we find the following second-quantized expression for the mixture's Hamiltonian:
\beqn\label{ham_mix_2nd}
& &  \hat H^{(ABC)} =
 \sum_{k,q} h^{(A)}_{kq} \hat \rho^{(A)}_{kq}
 + \frac{1}{2} \sum_{k,s,q,l} W^{(A)}_{ksql} \hat \rho^{(A)}_{kslq} 
 + \frac{1}{6}\sum_{k,s,p,r,l,q} U^{(A)}_{kspqlr} \hat \rho^{(A)}_{ksprlq} + \nonumber \\
& & + \sum_{k',q'} h^{(B)}_{k'q'} \hat \rho^{(B)}_{k'q'}
 + \frac{1}{2} \sum_{k',s',q',l'} W^{(B)}_{k's'q'l'} \hat \rho^{(B)}_{k's'l'q'} 
 + \frac{1}{6}\sum_{k',s',p',r',l',q'} U^{(B)}_{k's'p'q'l'r'} \hat \rho^{(B)}_{k's'p'r'l'q'} + \nonumber \\
& & + \sum_{k'',q''} h^{(C)}_{k''q''} \hat \rho^{(C)}_{k''q''}
 + \frac{1}{2} \sum_{k'',s'',q'',l''} W^{(C)}_{k''s''q''l''} \hat \rho^{(C)}_{k''s''l''q''} + \nonumber \\
& &
 + \frac{1}{6}\sum_{k'',s'',p'',r'',l'',q''} U^{(C)}_{k''s''p''q''l''r''} 
\hat \rho^{(C)}_{k''s''p''r''l''q''} + \nonumber \\
%%%%%%%%%%%% AB  AC  BC
& & + \sum_{k,k',q,q'} W^{(AB)}_{kk'qq'} \hat\rho^{(A)}_{kq} \hat\rho^{(B)}_{k'q'} 
    + \sum_{k,k'',q,q''} W^{(AC)}_{kk''qq''} \hat\rho^{(A)}_{kq} \hat\rho^{(C)}_{k''q''} 
    + \sum_{k',k'',q',q''} W^{(BC)}_{k'k''q'q''} \hat\rho^{(B)}_{k'q'} \hat\rho^{(C)}_{k''q''} + \nonumber \\
& & + \frac{1}{2} \sum_{k,k',s,q,q',l} U^{(AAB)}_{kk'sqq'l} \hat\rho^{(A)}_{kslq} \hat\rho^{(B)}_{k'q'}
    + \frac{1}{2} \sum_{k,k',s',q,q',l'} U^{(ABB)}_{kk's'qq'l'} \hat\rho^{(A)}_{kq} \hat\rho^{(B)}_{k's'l'q'} + 
\nonumber \\
& & + \frac{1}{2} \sum_{k,k'',s,q,q'',l} U^{(AAC)}_{kk''sqq''l} \hat\rho^{(A)}_{kslq} \hat\rho^{(C)}_{k''q''} 
    + \frac{1}{2} \sum_{k,k'',s'',q,q'',l''}  U^{(ACC)}_{kk''s''qq''l''} 
 \hat\rho^{(A)}_{kq} \hat\rho^{(C)}_{k''s''l''q''} + \nonumber \\
& & + \frac{1}{2} \sum_{k',k'',s',q',q'',l'} U^{(BBC)}_{k'k''s'q'q''l'} \hat\rho^{(B)}_{k's'l'q'} 
\hat\rho^{(C)}_{k''q''} 
+ \frac{1}{2} \sum_{k',k'',s'',q',q'',l''} U^{(BCC)}_{k'k''s''q'q''l''} 
\hat\rho^{(B)}_{k'q'} \hat\rho^{(C)}_{k''s''l''q''} + \nonumber \\
& & + \sum_{k,k',k'',q,q',q''} U^{(ABC)}_{kk'k''qq'q''} \hat\rho^{(A)}_{kq} \hat\rho^{(B)}_{k'q'}
\hat\rho^{(C)}_{k''q''}.  
\eeqn

$\hat H^{(ABC)}$
governs
the non-equilibrium dynamics (and statics)
of the mixture,
and the
most efficient way to treat
this dynamics is by
specifying the MCTDH method for the mixture,
making use of 
the building bricks
of the previous section \ref{SEC2}.
We see in (\ref{ham_mix_2nd})
two kinds of ingredients.
First, there are matrix elements (integrals)
of the various interaction terms with respect to the orbitals.
For the flow of exposition and for completeness,
we list them in Appendix \ref{appendix_C}.
Second,
there are various density operators in $\hat H^{(ABC)}$.
The $B$ and $C$ intra-species density operators 
can be read directly from Eqs.~(\ref{density_oper_1B},\ref{density_oper_2B_3B}),
when replacing therein the $A$-species quantities.
The inter-species density operators in (\ref{ham_mix_2nd})
can all be represented as 
appropriate products of
the one-body density operators:
$\left\{\hat \rho^{(A)}_{kq}\right\}$,
$\left\{\hat \rho^{(B)}_{k'q'}\right\}$
and 
$\left\{\hat \rho^{(C)}_{k''q''}\right\}$.
These are the (basic) building bricks
of our theory for mixtures.
But how to operate with them
on many-particle wave-functions of mixtures?

The multiconfigurational ansatz for a
mixture of three kinds of identical particles now 
takes on the from:
\beq\label{3Mix_ansatz}
\left|\Psi^{(ABC)}(t)\right> =
\sum^{N^{(A)}_{\mathit{conf}}}_{J_A=1}
\sum^{N^{(B)}_{\mathit{conf}}}_{J_B=1}
\sum^{N^{(C)}_{\mathit{conf}}}_{J_C=1}
C_{J_A,J_B,J_C}(t) \left|J_A,J_B,J_C;t\right>,
\eeq
where we denote hereafter $\vec J = (J_A, J_B, J_C)$ for brevity,
such that $C_{\vec J}(t) \equiv C_{J_A,J_B,J_C}(t)$,
$\left|\vec J;t\right> \equiv \left|J_A,J_B,J_C;t\right>$
and
$\sum_{\{\vec J\}} \equiv
\sum^{N^{(A)}_{\mathit{conf}}}_{J_A=1}
\sum^{N^{(B)}_{\mathit{conf}}}_{J_B=1}
\sum^{N^{(C)}_{\mathit{conf}}}_{J_C=1}$.

To prescribe the action of operators on the multiconfigurational
wave-function of the mixture (\ref{3Mix_ansatz}),
all we need to know is how the density operators 
operate on $\left|\Psi^{(ABC)}(t)\right>$.
The operation 
of the basic, one-body density operators,
whether $\hat \rho^{(A)}_{kq}$,
$\hat \rho^{(B)}_{k'q'}$
or
$\hat \rho^{(C)}_{k''q''}$
can be read of directly from Eqs.~(\ref{O_den}-\ref{C_three})
and we will not repeat them here 
(one needs just to replace therein $J_A$ by $\vec J$
in the overall notation,
and $M_A$ by $M_B$ or $M_C$,
when appropriate; also see \cite{mapping}).
For the inter-species two-body density operators we have:
\beq\label{2B_mix_dens_oper}
 C^{\hat \rho^{(A)}_{kq}\hat \rho^{(B)}_{k'q'}}_{\vec J}(t), \qquad
 C^{\hat \rho^{(A)}_{kq}\hat \rho^{(C)}_{k''q''}}_{\vec J}(t), \qquad 
 C^{\hat \rho^{(B)}_{k'q'}\hat \rho^{(C)}_{k''q''}}_{\vec J}(t).
\eeq
The notation in (\ref{2B_mix_dens_oper}) 
is to be understood as follows:
The two one-body density operators (in each case) are written 
as superscripts on the same level,
signifying that they commute one with the other;
The operation of the two one-body density operators
on $\left|\Psi^{(ABC)}(t)\right>$ is to be performed
sequentially, i.e., the first operates on
$\left|\Psi^{(ABC)}(t)\right>$ and the second 
operates on the outcome.
Finally,
for the inter-species three-body
density operators we have:
\beqn\label{3B_mix_dens_oper}
& &  
  C^{\hat \rho^{(A)}_{kslq} \hat \rho^{(B)}_{k'q'}}_{\vec J}(t), \qquad
  C^{\hat \rho^{(A)}_{kq} \hat \rho^{(B)}_{k's'l'q'}}_{\vec J}(t), \qquad
  C^{\hat \rho^{(A)}_{kslq} \hat \rho^{(C)}_{k''q''}}_{\vec J}(t), \nonumber \\
& &
  C^{\hat \rho^{(A)}_{kq} \hat \rho^{(C)}_{k''s''l''q''}}_{\vec J}(t), \qquad
  C^{\hat \rho^{(B)}_{k's'l'q'} \hat \rho^{(C)}_{k''q''}}_{\vec J}(t), \qquad
  C^{\hat \rho^{(B)}_{k'q'} \hat \rho^{(C)}_{k''s''l''q''}}_{\vec J}(t), \
\eeqn
where the operation of the
two-body density operators appearing in the superscripts 
is further decomposed to operations of one-body density operators
on $\left|\Psi^{(ABC)}(t)\right>$
analogously to Eq.~(\ref{basic_mapping_2B_3B}) [see Appendix \ref{appendix_A}], 
and
\beq\label{3B_mix_dens_oper_3}
C^{\hat \rho^{(A)}_{kq} \hat \rho^{(B)}_{k'q'} \hat \rho^{(C)}_{k''q''}}_{\vec J}(t).
\eeq
Now we are in the position to write the
action of operators on the multiconfigurational
wave-function of the mixture (\ref{3Mix_ansatz}).
This is collected for ease of reading and for completeness in
Appendix \ref{appendix_A}.

We have gathered most ingredients for the
derivation of the equations-of-motion,
which is written
down in the subsequence section \ref{SEC3.2}.
There are four possible mixtures (Fermi-Fermi-Fermi,
Bose-Fermi-Fermi, Bose-Bose-Fermi and Bose-Bose-Bose), 
and the resulting MCTDH-FFF,
MCTDH-BFF, MCTDH-BBF and MCTDH-BBB 
are 
to be derived 
and
presented in a unified manner,
in the spirit it has been done
in the single-species case \cite{book,unified} (and the previous section \ref{SEC2})
and for mixtures of two kinds of identical particles 
\cite{book,MCTDHX}.

\subsection{Equations-of-motion utilizing
one-body density operators and Combinadic-based mapping 
for mixtures}\label{SEC3.2} 

The action functional of the time-dependent
many-particle Schr\"odinger equation takes on the form:
\beqn\label{func_ABC}
 & &  S\left[\left\{C_{\vec J}(t)\right\},\left\{\phi_k(\x,t)\right\},
\left\{\psi_{k'}(\y,t)\right\},\left\{\chi_{k''}(\z,t)\right\}\right] = \\
 & & 
\int dt \Bigg\{\left< \Psi^{(ABC)}(t) \left| \hat H^{(ABC)} - i\frac{\partial}{\partial t}\right| \Psi^{(ABC)}(t)\right>
 - \nonumber \\
 & & - \sum_{k,j}^{M_A} \mu_{kj}^{(A)}(t) \left[\left<\phi_k \left|\right.\phi_j\right> - \delta_{kj}\right]
 - \sum_{k',j'}^{M_B} \mu_{k'j'}^{(B)}(t) \left[\left<\psi_{k'} \left|\right.\psi_{j'}\right> - \delta_{k'j'}\right]
- \nonumber \\
& & 
- \sum_{k'',j''}^{M_C} \mu_{k''j''}^{(C)}(t) \left[\left<\chi_{k''} \left|\right.\chi_{j''}\right> - 
\delta_{k''j''}\right]
- \varepsilon^{(ABC)}(t) \left[\sum_{\{\vec J\}} \left|C_{\vec J}(t)\right|^2 - 1 \right]\Bigg\}, \nonumber
\eeqn
where the time-dependent Lagrange multipliers 
$\left\{\mu_{kj}^{(A)}(t)\right\}$,
$\left\{\mu_{k'j'}^{(B)}(t)\right\}$
and\break\hfill 
$\left\{\mu_{k''j''}^{(C)}(t)\right\}$
are introduced, respectively, to ensure the orthonormality of the
$A$-, $B$- and $C$-species orbitals at all times.
Note that orbitals of distinct particles need
not be orthogonal to each other.
As for the single-species theory,
the Lagrange multiplier 
$\varepsilon^{(ABC)}(t)$
is redundant in the time-dependent case
and will resurface in the static theory.

In what follows we present the main steps of the derivation.
More details and various quantities needed
for the derivation and in particular for
the implementation of the equations-of-motion are deferred to Appendix \ref{appendix_B}
and Appendix \ref{appendix_C}.

To perform the variation of the action functional (\ref{func_ABC})
with respect to the coefficients,
we write the expectation value of $\hat H^{(ABC)}$ with respect to $\left|\Psi^{(ABC)}(t)\right>$
in a form which is 
explicit with respect to the coefficients:
\beqn\label{expectation_H_ABC_C}
& & \left<\Psi^{(ABC)}(t)\left| \hat H^{(ABC)} - i\frac{\partial}{\partial t} \right|\Psi^{(ABC)}(t)\right> = 
\nonumber \\ 
& & \qquad = \sum_{\{\vec J\}}
 C^\ast_{\vec J}(t) 
\left[ C^{\hat H^{(ABC)} - i\frac{\partial}{\partial t}^{(A)} - i\frac{\partial}{\partial t}^{(B)}
- i\frac{\partial}{\partial t}^{(C)}}_{\vec J}\!\!(t) - i \dot C_{\vec J}(t) \right]. 
\eeqn
The three time-derivative operators
$i\frac{\partial}{\partial t}^{(A)}$, $i\frac{\partial}{\partial t}^{(B)}$ and
$i\frac{\partial}{\partial t}^{(C)}$ make it clear that to each species
there is associated a different one-body operator 
representing the derivative of orbitals in time.

Performing the variation of 
$ S\left[\left\{C_{\vec J}(t)\right\},\left\{\phi_k(\x,t)\right\},
\left\{\psi_{k'}(\y,t)\right\},\left\{\chi_{k''}(\z,t)\right\}\right]$
with respect to the expansion coefficients
$\left\{ C^\ast_{\vec J}(t)\right\}$,
we then make use of the differential conditions for the orbitals of each species,
\beqn\label{diff_con_BC}
& & \left\{i\frac{\partial}{\partial t}^{(B)}\right\}_{k'q'} \equiv
 i\left<\psi_{k'} \left|\dot\psi_{q'}\right>\right. = 0, \ \ k',q'=1,\ldots,M_B, \nonumber \\
& & \left\{i\frac{\partial}{\partial t}^{(C)}\right\}_{k''q''} \equiv
 i\left<\chi_{k''} \left|\dot\chi_{q''}\right>\right. = 0, \ \ k'',q''=1,\ldots,M_C,
\eeqn
where the differential conditions with respect to
the $A$-spices orbitals have been introduced in (\ref{diff_con_A}).
This leads to the final result for the
equations-of-motion for the expansion 
coefficients:
\beqn\label{C_MIX_gen_phi_phidot}
& & C^{\hat H^{(ABC)}}_{\vec J}(t) = i \dot C_{\vec J}(t), \qquad \forall \vec J, \nonumber \\
& & C^{\hat H^{(ABC)}}_{\vec J}(t) =
 C^{\hat H^{(A)}}_{\vec J}(t) + C^{\hat H^{(B)}}_{\vec J}(t) + C^{\hat H^{(C)}}_{\vec J}(t) + \nonumber \\
& & + C^{\hat W^{(AB)}}_{\vec J}(t) + C^{\hat W^{(AC)}}_{\vec J}(t) + C^{\hat W^{(BC)}}_{\vec J}(t) + \nonumber \\
& & + C^{\hat U^{(AAB)}}_{\vec J}(t) + C^{\hat U^{(ABB)}}_{\vec J}(t) + C^{\hat U^{(AAC)}}_{\vec J}(t) +
      C^{\hat U^{(ACC)}}_{\vec J}(t) + \nonumber \\ 
& & + C^{\hat U^{(BBC)}}_{\vec J}(t) + C^{\hat U^{(BCC)}}_{\vec J}(t) + C^{\hat U^{(ABC)}}_{\vec J}(t). \
\eeqn
We remark that other forms of the 
differential conditions (\ref{diff_con_A},\ref{diff_con_BC})
can be used,
in particular,
each species can have a different form depending on the physical 
problem at hand and on numerical needs.

Let us 
move to the equations-of-motion
for the orbitals
$\left\{\phi_k(\x,t)\right\}$,
$\left\{\psi_{k'}(\y,t)\right\}$
and
$\left\{\chi_{k''}(\z,t)\right\}$.
For this,
we express the expectation value\break\hfill
$\left<\Psi^{(ABC)}\left|\hat H^{(ABC)} - i\frac{\partial}{\partial t}\right|\Psi^{(ABC)}\right>$
in a form which explicitly depends 
on the various
integrals with respect to the orbitals.
The result is lengthly and posted
in Appendix \ref{appendix_C}.
In particular,
the expectation values of the various density
operators in $\hat H^{(ABC)}$ [Eq.~(\ref{ham_mix_2nd})]
emerge 
as matrix elements of
the different intra-species and inter-species
reduced density matrices.
For ease of reading and for completeness,
we collect in Appendix \ref{appendix_B}
all reduced density matrices
and their respective matrix elements needed in the theory
and its numerical implementation.

We can now proceed and perform the variation
of the action functional (\ref{func_ABC}) with respect to the orbitals.
Performing the variation
with respect to 
$\left\{\phi^\ast_k(\x,t)\right\}$,
$\left\{\psi^\ast_{k'}(\y,t)\right\}$
and
$\left\{\chi^\ast_{k''}(\z,t)\right\}$,
making use of the orthonormality relations
of each species' orbitals,
we solve for the Lagrange multipliers,
$k,j=1,\ldots,M_A$,
$k',j'=1,\ldots,M_B$
and
$k'',j''=1,\ldots,M_C$:
\beqn\label{LM_A_B_C}
& & \mu_{kj}^{(A)}(t) = 
\left<\phi_j\left| 
\sum^{M_A}_{q=1} \left( \rho^{(A)}_{kq} \left[ \hat h^{(A)} 
 - i\frac{\partial}{\partial t}^{(A)} \right] + 
\{\rho_2 \hat W\}^{(A)}_{kq} +  \{\rho_3 \hat U\}^{(A)}_{kq}
\right) \right|\phi_q\right>, \ \\
 & & \mu_{k'j'}^{(B)}(t) = 
\left<\psi_{j'}\left| 
\sum^{M_B}_{q'=1} \left( \rho^{(B)}_{k'q'} \left[ \hat h^{(B)} 
 - i\frac{\partial}{\partial t}^{(B)} \right] + 
\{\rho_2 \hat W\}^{(B)}_{k'q'} +  \{\rho_3 \hat U\}^{(B)}_{k'q'}
\right) \right|\psi_{q'}\right>, \nonumber \\
 & & \mu_{k''j''}^{(C)}(t) = 
\left<\chi_{j''}\left| 
\sum^{M_C}_{q''=1} \left( \rho^{(C)}_{k''q''} \left[ \hat h^{(C)} 
 - i\frac{\partial}{\partial t}^{(C)} \right] + 
\{\rho_2 \hat W\}^{(C)}_{k''q''} +  \{\rho_3 \hat U\}^{(C)}_{k''q''}
\right) \right|\chi_{q''}\right>. \nonumber \
\eeqn
The terms appearing in the Lagrange
multipliers are all defined in Appendix \ref{appendix_C}.
We discuss them below,
after we arrive at the final form of the equations-of-motion 
for the orbitals.

%%%%%%%%%%%%%%%%%%%%% PROJECTORS FOR MIXTURES 
To proceed we introduce 
the projection operators for the mixture:
\beq\label{project_BC}
\hat {\mathbf P}^{(B)} = 1 - \sum_{u'=1}^{M_B} \left|\psi_{u'}\right>\left<\psi_{u'}\right|, \qquad
\hat {\mathbf P}^{(C)} = 1 - \sum_{u''=1}^{M_C} \left|\chi_{u''}\right>\left<\chi_{u''}\right|,
\eeq
where the projection operator
for the $A$-species orbitals $\hat {\mathbf P}^{(A)}$ 
was defined in (\ref{project_A}). 
Now,
eliminating the Lagrange multipliers (\ref{LM_A_B_C}) 
and making
use of the differential conditions 
for each species (\ref{diff_con_A},\ref{diff_con_BC}), 
we
obtain the final form
of the equations-of-motion
of the orbitals of the mixture,
$j=1,\ldots,M_A$,
$j'=1,\ldots,M_B$ and
$j''=1,\ldots,M_C$:
\beqn\label{EOM_final_orbitals_3mix}
& & \!\!\!\!\!\!\!\! i\left|\dot\phi_j\right> =  \hat {\mathbf P}^{(A)} 
\left[\hat h^{(A)} \left|\phi_j\right>  +
\sum^{M_A}_{k,q=1}  \left\{\brho^{(A)}(t)\right\}^{-1}_{jk}
 \bigg( \{\rho_2 \hat W\}^{(A)}_{kq} +  \{\rho_3 \hat U\}^{(A)}_{kq} \bigg)
\left|\phi_q\right> \right], \\
& & \!\!\!\!\!\!\!\! i\left|\dot\psi_{j'}\right> =  \hat {\mathbf P}^{(B)} 
\left[\hat h^{(B)} \left|\psi_{j'}\right>  +
\sum^{M_B}_{k',q'=1}  \left\{\brho^{(B)}(t)\right\}^{-1}_{j'k'}
 \bigg( \{\rho_2 \hat W\}^{(B)}_{k'q'} +  \{\rho_3 \hat U\}^{(B)}_{k'q'} \bigg)
\left|\psi_{q'}\right> \right], \nonumber \\
& & \!\!\!\!\!\!\!\! i\left|\dot\chi_{j''}\right> =  \hat {\mathbf P}^{(C)} 
\left[\hat h^{(C)} \left|\chi_{j''}\right>  +
\sum^{M_C}_{k'',q''=1}  \left\{\brho^{(C)}(t)\right\}^{-1}_{j''k''}
 \bigg( \{\rho_2 \hat W\}^{(C)}_{k''q''} + \{\rho_3 \hat U\}^{(C)}_{k''q''} \bigg)
\left|\chi_{q''}\right> \right]. \nonumber \
\eeqn
We see the appealing structure of the equations-of-motion
for the orbitals.
The various one-body operators which assemble the contributions from different orders of the 
interactions, corresponding to the one-, two- and three-body parts of
the many-particle Hamiltonian $\hat H^{(ABC)}$ (\ref{ham_3mix_general}),
are separated.
Moreover,
it is seen that each one-body operator is comprised 
of
products of reduced density matrices of increasing order
times one-body potentials resulting from
interactions of the same order (see Appendix \ref{appendix_C}
for the explicit terms).
This separation, 
originally put forward for the
first time
in this context for the single-species static theory for bosons MCHB \cite{MCHB},
is not 
only theoretically appealing,
but is expected to  
make the implementation of the theory 
in case 
of higher-body 
forces 
further
efficient.

Equations-of-motion (\ref{EOM_final_orbitals_3mix}) for the
orbitals together with (\ref{C_MIX_gen_phi_phidot}) 
for the expansion coefficients constitute 
the
propagation theory for mixtures of three kinds
of identical particles,
interacting with all possible interactions
up to three-body forces.
All four possible mixtures (Fermi-Fermi-Fermi,
Bose-Fermi-Fermi, Bose-Bose-Fermi and Bose-Bose-Bose) 
are presented in a unified manner,
the respective acronyms are denoted 
as MCTDH-FFF, MCTDH-BFF, MCTDH-BBF and
MCTDH-BBB.

To conclude our work,
we note that one 
can compute with imaginary time propagation
for time-independent
Hamiltonians self-consistent ground 
and excited states for 
3-species mixtures.
Substituting
$t \to -it$ into the equations-of-motion 
for the coefficients and orbitals, Eqs.~(\ref{C_MIX_gen_phi_phidot},\ref{EOM_final_orbitals_3mix}),
the final time-independent (static) 
theory reads,
$k=1,\ldots,M_A$,
$k'=1,\ldots,M_B$ and
$k''=1,\ldots,M_C$:
\beqn\label{statical_3mix}
 & & 
\sum_{q=1}^{M_A} \left[ \rho^{(A)}_{kq} \hat h^{(A)} +
 \{\rho_2 \hat W\}^{(A)}_{kq} +  \{\rho_3 \hat U\}^{(A)}_{kq} 
\right] \left|\phi_q\right> =
 \sum_{j=1}^{M_A} \mu_{kj}^{(A)} \left|\phi_j\right>, \nonumber \\
 & & 
\sum_{q'=1}^{M_B} \left[ \rho^{(B)}_{k'q'} \hat h^{(B)} +
 \{\rho_2 \hat W\}^{(B)}_{k'q'} +  \{\rho_3 \hat U\}^{(B)}_{k'q'} 
\right] \left|\psi_{q'}\right> =
 \sum_{j'=1}^{M_B} \mu_{k'j'}^{(B)} \left|\psi_{j'}\right>, \nonumber \\
 & & 
\sum_{q''=1}^{M_C} \left[ \rho^{(C)}_{k''q''} \hat h^{(C)} +
 \{\rho_2 \hat W\}^{(C)}_{k''q''} +  \{\rho_3 \hat U\}^{(C)}_{k''q''} 
\right] \left|\chi_{q''}\right> =
 \sum_{j''=1}^{M_C} \mu_{k''j''}^{(C)} \left|\chi_{j''}\right>, \nonumber \\
 & & 
\qquad \qquad 
C^{\hat H^{(ABC)}}_{\vec J} = \varepsilon^{(ABC)} C_{\vec J}, \qquad \forall \vec J, \
\eeqn
where, making use of the normalization of the static 
many-particle wave-function $\left|\Psi^{(ABC)}\right>$, 
$\varepsilon^{(ABC)}= \sum_{\vec J} C^\ast_{\vec J} C^{\hat H^{(ABC)}}_{\vec J}$
is the eigen-energy of the system.
Finally,
utilizing 
the fact that the matrices of Lagrange multipliers
$\{\mu_{kj}^{(A)}\}$, $\{\mu_{k'j'}^{(B)}\}$ and $\{\mu_{k''j''}^{(C)}\}$   
are Hermitian (for stationary states)
and of the 
invariance property of the multiconfigurational wave-function
(to unitary transformations of each species' 
orbitals compensated by the `reverse' transformations
of the coefficients), 
one can transform Eq.~(\ref{statical_3mix})
to a representation where
$\{\mu_{kj}^{(A)}\}$, $\{\mu_{k'j'}^{(B)}\}$ and $\{\mu_{k''j''}^{(C)}\}$
are diagonal matrices.
This concludes our 
derivations.

\section{Brief summary and outlook}\label{SEC4}

In the present work we have specified the MCTDH
method for a new complicated system of relevance.
We have considered mixtures
of three kinds of identical particles
interacting via all combinations of two- and three-body forces.
We have derived the equations-of-motion for
the expansion coefficients, $\left\{C_{\vec J}(t)\right\}$,
and 
the orbitals, 
$\left\{\phi_k(\x,t)\right\}$,
$\left\{\psi_{k'}(\y,t)\right\}$
and 
$\left\{\chi_{k''}(\z,t)\right\}$,
see Eqs.~(\ref{C_MIX_gen_phi_phidot},\ref{EOM_final_orbitals_3mix}).
The self-consistent
static theory has 
been derived as well,
see Eq.~(\ref{statical_3mix}).

All quantities needed for the implementation
of the theory have been prescribed in details.
On the methodological level,
we have represented the 
coefficients' part of the equations-of-motion
in a compact recursive form 
in terms of one-body density operators only,
$\left\{\hat \rho^{(A)}_{kq}\right\}$,
$\left\{\hat \rho^{(B)}_{k'q'}\right\}$
and
$\left\{\hat \rho^{(C)}_{k''q''}\right\}$.
The recursion utilizes the recently proposed 
Combinadic-based mapping for fermionic and bosonic operators in Fock space \cite{mapping}
that has been 
successfully applied and implemented within the MCTDHB package \cite{package}.
Our derivation sheds
new light on the
representation of the coefficients'
part in MCTDHF and MCTDHB 
without resorting to the 
matrix elements of the many-body Hamiltonian 
with respect to the time-dependent configurations,
and suggests a recipe for 
efficient implementation of 
MCTDH-FFF, MCTDH-BFF, MCTDH-BBF and
MCTDH-BBB 
which is well-suitable 
for parallel implementation.

As an outlook of the present theory,
let us imagine the possibility of conversion between the distinct particles,
say the conversion of the $A$ and $B$ species
to the $C$ species,
which can be written symbolically as the following ``reaction'':
$$
A + B \leftrightharpoons C.
$$
Such a process would be a model, e.g.,
for the resonant association
of hetero-nuclear ultra-cold molecules.
The derivation of an efficient MCTDH-{\it conversion} theory
in this case 
would require the extension of the 
Combinadic-based
mapping \cite{mapping} 
to systems with particle conversion,
and the assembly of more building bricks
than just the one-body density operators used in the 
present theory,
$\left\{\hat \rho^{(A)}_{kq}\right\}$,
$\left\{\hat \rho^{(B)}_{k'q'}\right\}$
and
$\left\{\hat \rho^{(C)}_{k''q''}\right\}$.

\section*{Acknowledgments}

The paper is dedicated 
to Professor Debashis Mukherjee, 
a dear colleague and friend,
on the occasion of his 65{\it th} birthday.
We are grateful to Hans-Dieter Meyer for multiple and continuous discussions on MCTDH,
and acknowledge financial support by the DFG.

\appendix

\section{Calculating expectation values 
of operators in mixtures of three kinds of identical particles}\label{appendix_A}

Following \cite{mapping},
we write the general expectation value of an operator 
$\hat O^{(3mix)}$ in a 3-species mixture as follows:
\beqn\label{expectation_3}
& & \left<\Psi^{(ABC)}(t)\left| \hat O^{(3mix)} \right|\Psi^{(ABC)}(t)\right> =  
\left<\Psi^{(ABC)}(t)\left| \left\{ \hat O^{(3mix)} \right|\Psi^{(ABC)}(t)\right> \right\} = \nonumber \\
& & = \sum_{{\vec J}} C^\ast_{\vec J}(t) C^{\hat O^{(3mix)}}_{\vec J}(t), 
\eeqn
where
\beq\label{O_Psi_3}
 \hat O^{(3mix)} \left|\Psi^{(ABC)}(t)\right> = 
\hat O^{(3mix)} \sum_{{\vec J}} C_{\vec J}(t) \left|\vec J;t\right> \equiv 
  \sum_{{\vec J}} C^{\hat O^{(3mix)}}_{\vec J}(t) \left|\vec J;t\right>.
\eeq
$\hat O^{(3mix)}$ can be a
one-, two- or three-body operator or
any combination thereof.

The operation of single-species operators,
whether $\hat O^{(A)}$, $\hat O^{(B)}$ 
or
$\hat O^{(C)}$ can be read of directly from Eqs.~(\ref{O_den}-\ref{C_three})
and we will not repeat them here 
(one needs just to replace therein $J_A$ by $\vec J$
in the overall notation,
and $M_A$ by $M_B$ or $M_C$,
when appropriate; also see \cite{mapping}).

For the inter-species two-body operators we prescribe
the compact result for completeness.
For the two-body operators
$\hat O^{(AB)} = \sum_{k,k',q,q'} O^{(AB)}_{kk'qq'} \hat \rho^{(A)}_{kq} \hat \rho^{(B)}_{k'q'}$,\break\hfill
$\hat O^{(AC)} = \sum_{k,k'',q,q''} O^{(AC)}_{kk''qq''} \hat \rho^{(A)}_{kq} \hat \rho^{(C)}_{k''q''}$
and 
$\hat O^{(BC)} = \sum_{k',k'',q',q''} O^{(BC)}_{k'k''q'q''} \hat \rho^{(B)}_{k'q'} \hat \rho^{(C)}_{k''q''}$
we find:
\beqn\label{C_3mix_2B}
 C^{\hat O^{(AB)}}_{\vec J}(t) &=& \sum_{k,k',q,q'=1}^{M_A,M_B} O^{(AB)}_{kk'qq'} 
  C^{\hat \rho^{(A)}_{kq}\hat \rho^{(B)}_{k'q'}}_{\vec J}(t), \nonumber \\
 C^{\hat O^{(AC)}}_{\vec J}(t) &=& \sum_{k,k'',q,q''=1}^{M_A,M_C} O^{(AC)}_{kk''qq''} 
  C^{\hat \rho^{(A)}_{kq}\hat \rho^{(C)}_{k''q''}}_{\vec J}(t), \nonumber \\
 C^{\hat O^{(BC)}}_{\vec J}(t) &=& \sum_{k',k'',q',q''=1}^{M_B,M_C} O^{(BC)}_{k'k''q'q''} 
  C^{\hat \rho^{(B)}_{k'q'}\hat \rho^{(C)}_{k''q''}}_{\vec J}(t). \
\eeqn
Note the factorization
of the one-body (basic) density operators for
the inter-species operators,
which simplify the way 
how the coefficients' vector is evaluated.

For the inter-species three-body operators 
resulting from the force between two identical particles and a third distinct one
we list the final result for completeness.
For the three-body operators
\beqn\label{3B_operators}
\hat O^{(AAB)} &=& \frac{1}{2} \sum_{k,k',s,q,q'l} O^{(AAB)}_{kk'sqq'l} \hat \rho^{(A)}_{kslq} \hat \rho^{(B)}_{k'q'},
\nonumber \\
\hat O^{(ABB)} &=& \frac{1}{2} \sum_{k,k',s',q,q',l'} O^{(ABB)}_{kk's'qq'l'} \hat \rho^{(A)}_{kq} 
\hat \rho^{(B)}_{k's'l'q'}, \nonumber \\
\hat O^{(AAC)} &=& \frac{1}{2} \sum_{k,k'',s,q,q''l} O^{(AAC)}_{kk''sqq''l} \hat \rho^{(A)}_{kslq} 
\hat \rho^{(C)}_{k''q''}, \nonumber \\
\hat O^{(ACC)} &=& \frac{1}{2} \sum_{k,k'',s'',q,q'',l''} O^{(ACC)}_{kk''s''qq''l''} \hat \rho^{(A)}_{kq} 
\hat \rho^{(C)}_{k''s''l''q''}, \nonumber \\
\hat O^{(BBC)} &=& \frac{1}{2} \sum_{k',k'',s',q',q''l'} O^{(BBC)}_{k'k''s'q'q''l'} \hat \rho^{(B)}_{k's'l'q'} 
\hat \rho^{(C)}_{k''q''}, \nonumber \\
\hat O^{(BCC)} &=& \frac{1}{2} \sum_{k',k'',s'',q',q'',l''} O^{(BCC)}_{k'k''s''q'q''l''} \hat \rho^{(B)}_{k'q'} 
\hat \rho^{(C)}_{k''s''l''q''}, \
\eeqn
we find:
\beqn\label{C_3mix_binary_3B}
& & C^{\hat O^{(AAB)}}_{\vec J}(t) = \nonumber \\
&=& \frac{1}{2} \sum_{k,k',s,q,q',l=1}^{M_A,M_B} O^{(AAB)}_{kk'sqq'l} 
\left[ \pm \delta_{sl} C^{\hat \rho^{(A)}_{kq}\hat \rho^{(B)}_{k'q'}}_{\vec J}(t) 
\mp {C^{\hat \rho^{(A)}_{sq}\hat \rho^{(B)}_{k'q'}}_{\vec J}}^{\hat \rho^{(A)}_{kl}}\!(t) \right], 
\nonumber \\ 
& & C^{\hat O^{(ABB)}}_{\vec J}(t) = \nonumber \\ 
&=& \frac{1}{2} \sum_{k,k',s',q,q',l'=1}^{M_A,M_B} O^{(ABB)}_{kk's'qq'l'} \left[ \pm \delta_{s'l'} 
C^{\hat \rho^{(A)}_{kq} \hat \rho^{(B)}_{k'q'}}_{\vec J}(t) 
\mp {C^{\hat \rho^{(A)}_{kq} \hat \rho^{(B)}_{s'q'}}_{\vec J}}^{\hat \rho^{(B)}_{k'l'}}\!(t) \right],
\nonumber \\
& & C^{\hat O^{(AAC)}}_{\vec J}(t) = \nonumber \\ 
&=& \frac{1}{2} \sum_{k,k'',s,q,q'',l=1}^{M_A,M_C} O^{(AAC)}_{kk''sqq''l} 
\left[ \pm \delta_{sl} C^{\hat \rho^{(A)}_{kq}\hat \rho^{(C)}_{k''q''}}_{\vec J}(t) 
\mp {C^{\hat \rho^{(A)}_{sq}\hat \rho^{(C)}_{k''q''}}_{\vec J}}^{\hat \rho^{(A)}_{kl}}\!(t) \right],
\ \\ 
& & C^{\hat O^{(ACC)}}_{\vec J}(t) = \nonumber \\
&=& \frac{1}{2} \sum_{k,k'',s'',q,q'',l''=1}^{M_A,M_C} O^{(ACC)}_{kk''s''qq''l''} \left[ \pm \delta_{s''l''} 
C^{\hat \rho^{(A)}_{kq} \hat \rho^{(C)}_{k''q''}}_{\vec J}(t) 
\mp {C^{\hat \rho^{(A)}_{kq} \hat \rho^{(C)}_{s''q''}}_{\vec J}}^{\hat \rho^{(C)}_{k''l''}}\!(t) \right], 
\nonumber \\
 & & C^{\hat O^{(BBC)}}_{\vec J}(t) = \nonumber \\
 &=& \frac{1}{2} \sum_{k',k'',s',q',q'',l'=1}^{M_B,M_C} O^{(BBC)}_{k'k''s'q'q''l'} 
\left[ \pm \delta_{s'l'} C^{\hat \rho^{(B)}_{k'q'}\hat \rho^{(C)}_{k''q''}}_{\vec J}(t) 
\mp {C^{\hat \rho^{(B)}_{s'q'}\hat \rho^{(C)}_{k''q''}}_{\vec J}}^{\hat \rho^{(B)}_{k'l'}}\!(t) \right],
\nonumber \\ 
& & C^{\hat O^{(BCC)}}_{\vec J}(t) = \nonumber \\
&=& \frac{1}{2} \sum_{k',k'',s'',q',q'',l''=1}^{M_B,M_C} O^{(BCC)}_{k'k''s''q'q''l''} \left[ \pm \delta_{s''l''} 
C^{\hat \rho^{(B)}_{k'q'} \hat \rho^{(C)}_{k''q''}}_{\vec J}(t) 
\mp {C^{\hat \rho^{(B)}_{k'q'} \hat \rho^{(C)}_{s''q''}}_{\vec J}}^{\hat \rho^{(C)}_{k''l''}}\!(t) \right].
\nonumber 
\eeqn
We remind that the appearance
of the one-body (basic) density operators 
on two levels means that the lower-level
multiplication has to be performed first,
and the upper-level second.

Finally, 
for the inter-species three-body operator we
give the closed-form result for completeness.
For the three-body operator\break\hfill
$\hat O^{(ABC)} = \sum_{k,k',k'',q,q',q''} O^{(ABC)}_{kk'k''qq'q''} 
\hat \rho^{(A)}_{kq} \hat \rho^{(B)}_{k'q'} \hat \rho^{(C)}_{k''q''}$
we find:
\beq\label{C_3mix_3B}
 C^{\hat O^{(ABC)}}_{\vec J}(t) = \sum_{k,k',k'',q,q',q''=1}^{M_A,M_B,M_C} O^{(ABC)}_{kk'k''qq'q''} 
  C^{\hat \rho^{(A)}_{kq} \hat \rho^{(B)}_{k'q'} \hat \rho^{(C)}_{k''q''}}_{\vec J}(t),
\eeq
which concludes our Combinadic-based \cite{mapping} 
representation of the equations-of-motion
for the coefficients 
in MCTDH for mixtures
of 3 kinds of identical particles interacting with up to 3-body forces,
and the calculations of all relevant matrix elements
with respect to $\left|\Psi^{(ABC)}(t)\right>$.

\section{Reduced density matrices for mixtures of three kinds
of identical particles interacting with up to three-body forces}\label{appendix_B}

\subsection*{Intra-species reduced density matrices}

The reduced one-body density matrix of the single-species multiconfigurational
wave-function $\left|\Psi^{(A)}(t)\right>$ is given by:
\beqn\label{DNS_A_1}
 & & \rho^{(A)}(\x_1|\x'_1;t) = N_A \int d\x_2 d\x_3 \cdots d\x_{N_A} \\
& & = {\Psi^{(A)}}^\ast(\x'_1,\x_2,\ldots,\x_{N_A};t)
 \Psi^{(A)}(\x_1,\x_2,\ldots,\x_{N_A};t)  = \nonumber \\
  & & = \left<\Psi^{(A)}(t)\left|\left\{\hat{\mathbf \Psi}_A^\dag(\x'_1)\hat{\mathbf \Psi}_A(\x_1)
\right|\Psi^{(A)}(t)\right>\right\} =
 \sum^M_{k,q=1} \rho^{(A)}_{kq}(t) \phi^\ast_k(\x'_1,t)\phi_q(\x_1,t), \nonumber \
\eeqn
where its matrix elements in the orbital basis 
$\rho^{(A)}_{kq}(t)$
are given in Eq.~(\ref{denisty_matrx_element}) of the main text.

Then,
the reduced two-body density matrix of the single-species multiconfigurational
wave-function $\left|\Psi^{(A)}(t)\right>$ is given by:
\beqn\label{DNS_A_2}
& &  \rho^{(A)}(\x_1,\x_2|\x'_1,\x'_2;t) = N_A(N_A-1) \int d\x_3 \cdots d\x_{N_A} \times \\
& & \times  {\Psi^{(A)}}^\ast(\x'_1,\x'_2,\x_3,\ldots,\x_{N_A};t)  \Psi^{(A)}(\x_1,\x_2,\x_3,\ldots,\x_{N_A};t)
     = \nonumber \\
& & = \left<\Psi^{(A)}(t)\left|\left\{\hat{\mathbf \Psi}_A^\dag(\x'_1)\hat{\mathbf \Psi}_A^\dag(\x'_2)
\hat{\mathbf \Psi}_A(\x_2)\hat{\mathbf \Psi}_A(\x_1)\right|\Psi^{(A)}(t)\right> \right\} = \nonumber \\
 & & = \sum^M_{k,s,l,q=1} \rho^{(A)}_{kslq}(t)
\phi^\ast_k(\x'_1,t) \phi^\ast_s(\x'_2,t) \phi_l(\x_2,t) \phi_q(\x_1,t), \nonumber \
\eeqn
where its matrix elements in the orbital basis 
$\rho^{(A)}_{kslq}(t)$
are given in Eq.~(\ref{denisty_matrx_element}).

Finally in the single-species case,
the reduced three-body density matrix of $\left|\Psi^{(A)}(t)\right>$ is given by:
\beqn\label{DNS_A_3}
& & \rho^{(A)}(\x_1,\x_2,\x_3|\x'_1,\x'_2,\x'_3;t) = N_A(N_A-1)(N_A-2) \int d\x_4 \cdots d\x_{N_A} \times \nonumber \\
 & & \times  {\Psi^{(A)}}^\ast(\x'_1,\x'_2,\x'_3,\x_4,\ldots,\x_{N_A};t)  \Psi^{(A)} (\x_1,\x_2,\x_3,\x_4,\ldots,\x_{N_A};t)
     =  \\
& & = \left<\Psi^{(A)}(t)\left|\left\{\hat{\mathbf \Psi}_A^\dag(\x'_1)\hat{\mathbf \Psi}_A^\dag(\x'_2)
\hat{\mathbf \Psi}_A^\dag(\x'_3)
 \hat{\mathbf \Psi}_A(\x_3)\hat{\mathbf \Psi}_A(\x_2)
\hat{\mathbf \Psi}_A(\x_1)\right|\Psi^{(A)}(t)\right>\right\} = \nonumber \\
 & & = \sum^M_{k,s,p,r,l,q=1} \rho^{(A)}_{ksprlq}(t)
\phi^\ast_k(\x'_1,t) \phi^\ast_s(\x'_2,t) \phi^\ast_p(\x'_3,t) \phi_r(\x_3,t) \phi_l(\x_2,t) \phi_q(\x_1,t), \nonumber \
\eeqn
where its matrix elements in the orbital basis
$\rho^{(A)}_{ksprlq}(t)$
are given in Eq.~(\ref{denisty_matrx_element}).

The reduced density matrices of the $B$ and $C$ species
are defined in an analogous manner,
where $B$ and $C$ quantities are to replace
the $A$ quantities in Eqs.~(\ref{DNS_A_1}-\ref{DNS_A_3}).

\subsection*{Inter-species reduced two-body density matrices}

For completeness,
we give all inter-species
reduced density matrices that
occur in a mixture 
of three kinds of identical particles
interacting
with 
up to
three-body forces,
where each species may have a different spin.
There are three such reduced density matrices
which are associated with
the two-body interactions 
of two
distinct particles.

\beqn\label{DNS_AB}
& & \rho^{(AB)}(\x_1,\y_1|\x'_1,\y'_1;t) = N_A N_B \int \, d\x_2 \cdots d\x_{N_A} 
d\y_2 \cdots d\y_{N_B} d\z_1 \cdots d\z_{N_C} \times \nonumber \\
& & \times {\Psi^{(ABC)}}^\ast(\x'_1,\ldots,\x_{N_A},\y'_1,\ldots,\y_{N_B},\z_1,\ldots,\z_{N_C};t) \times \nonumber \\
& & \times \Psi^{(ABC)}(\x_1,\ldots,\x_{N_A},\y_1,\ldots,\y_{N_B},\z_1,\ldots,\z_{N_C};t) = \nonumber \\
& & = \left<\Psi^{(ABC)}(t)\left| \left\{ 
\hat{\mathbf \Psi}_A^\dag(\x'_1)
\hat{\mathbf \Psi}_A(\x_1)
\hat{\mathbf \Psi}_B^\dag(\y'_1)
\hat{\mathbf \Psi}_B(\y_1)\right|\Psi^{(ABC)}(t)\right> \right\} = \nonumber \\
& & = \sum^{M_A,M_B}_{k,k',q,q'=1} 
\rho^{(AB)}_{kk'qq'}(t) \phi^\ast_{k}(\x'_1,t) \phi_{q}(\x_1,t)
\psi^\ast_{k'}(\y'_1,t) \psi_{q'}(\y_1,t), \
\eeqn
where its matrix elements in the orbital basis are give by:
\beq
\rho^{(AB)}_{kk'qq'}(t)=
\sum_{\vec J} C^\ast_{\vec J}(t) C^{\hat \rho^{(A)}_{kq}\hat \rho^{(B)}_{k'q'}}_{\vec J}(t).
\eeq

\beqn\label{DNS_AC}
& & \rho^{(AC)}(\x_1,\z_1|\x'_1,\z'_1;t) = N_A N_C \int d\x_2 \cdots d\x_{N_A} 
d\y_1 \cdots d\y_{N_B} d\z_2 \cdots d\z_{N_C} \times \nonumber \\
& & \times {\Psi^{(ABC)}}^\ast(\x'_1,\ldots,\x_{N_A},\y_1,\ldots,\y_{N_B},\z'_1,\ldots,\z_{N_C};t) \times \nonumber \\
& & \times \Psi^{(ABC)}(\x_1,\ldots,\x_{N_A},\y_1,\ldots,\y_{N_B},\z_1,\ldots,\z_{N_C};t) = \nonumber \\
& & = \left<\Psi^{(ABC)}(t)\left| \left\{
\hat{\mathbf \Psi}_A^\dag(\x'_1)
\hat{\mathbf \Psi}_A(\x_1)
\hat{\mathbf \Psi}_C^\dag(\z'_1)
\hat{\mathbf \Psi}_C(\z_1)\right|\Psi^{(ABC)}(t)\right> \right\} = \nonumber \\
& & = \sum^{M_A,M_B}_{k,k'',q,q''=1} 
\rho^{(AC)}_{kk''qq''}(t) \phi^\ast_{k}(\x'_1,t) \phi_{q}(\x_1,t)
\chi^\ast_{k''}(\z'_1,t) \chi_{q''}(\z_1,t), \
\eeqn
where its matrix elements in the orbital basis are given by:
\beq
\rho^{(AC)}_{kk''qq''}(t)=
\sum_{\vec J} C^\ast_{\vec J}(t) C^{\hat \rho^{(A)}_{kq}\hat \rho^{(C)}_{k''q''}}_{\vec J}(t).
\eeq

\beqn\label{DNS_BC}
& & \rho^{(BC)}(\y_1,\z_1|\y'_1,\z'_1;t) = N_B N_C \int d\x_1 \cdots d\x_{N_A} 
d\y_2 \cdots d\y_{N_B} d\z_2 \cdots d\z_{N_C} \times \nonumber \\
& & \times {\Psi^{(ABC)}}^\ast(\x_1,\ldots,\x_{N_A},\y'_1,\ldots,\y_{N_B},\z'_1,\ldots,\z_{N_C};t) \times \nonumber \\
& & \times \Psi^{(ABC)}(\x_1,\ldots,\x_{N_A},\y_1,\ldots,\y_{N_B},\z_1,\ldots,\z_{N_C};t) = \nonumber \\
& & = \left<\Psi^{(ABC)}(t)\left| \left\{
\hat{\mathbf \Psi}_B^\dag(\y'_1)
\hat{\mathbf \Psi}_B(\y_1)
\hat{\mathbf \Psi}_C^\dag(\z'_1)
\hat{\mathbf \Psi}_C(\z_1)\right|\Psi^{(ABC)}(t)\right> \right\} = \nonumber \\
& & = \sum^{M_B,M_B}_{k',k'',q',q''=1} 
\rho^{(BC)}_{k'k''q'q''}(t) \psi^\ast_{k'}(\y'_1,t) \psi_{q'}(\y_1,t)
\chi^\ast_{k''}(\z'_1,t) \chi_{q''}(\z_1,t), \
\eeqn
where its matrix elements in the orbital basis are given by:
\beq
\rho^{(AC)}_{k'k''q'q''}(t)=
\sum_{\vec J} C^\ast_{\vec J}(t) C^{\hat \rho^{(B)}_{k'q'}\hat \rho^{(C)}_{k''q''}}_{\vec J}(t).
\eeq

\subsection*{Inter-species reduced three-body density matrices}

There are six reduced three-body density matrices
which are associated with
the three-body interactions of two
identical particles with a third distinct one.

\beqn\label{DNS_AAB}
& & \rho^{(AAB)}(\x_1,\x_2,\y_1|\x'_1,\x'_2,\y'_1;t) = \\
 & & = N_A(N_A-1) N_B \int \, d\x_3 \cdots d\x_{N_A} 
d\y_2 \cdots d\y_{N_B} d\z_1 \cdots d\z_{N_C} \times \nonumber \\
& & \times {\Psi^{(ABC)}}^\ast(\x'_1,\x'_2,\ldots,\x_{N_A},\y'_1,\y_2,\ldots,\y_{N_B},\z_1,\z_2,\ldots,\z_{N_C};t) 
\times \nonumber \\
& & \times \Psi^{(ABC)}(\x_1,\x_2,\ldots,\x_{N_A},\y_1,\y_2,\ldots,\y_{N_B},\z_1,\z_2,\ldots,\z_{N_C};t) = \nonumber \\
& & = \left<\Psi^{(ABC)}(t)\left| \left\{ 
\hat{\mathbf \Psi}_A^\dag(\x'_1)
\hat{\mathbf \Psi}_A^\dag(\x'_2)
\hat{\mathbf \Psi}_A(\x_2)
\hat{\mathbf \Psi}_A(\x_1)
\hat{\mathbf \Psi}_B^\dag(\y'_1)
\hat{\mathbf \Psi}_B(\y_1)\right|\Psi^{(ABC)}(t)\right> \right\} = \nonumber \\
& & = \sum^{M_A,M_B}_{k,k',s,l,q,q'=1} 
\rho^{(AAB)}_{kk'slqq'}(t) \phi^\ast_{k}(\x'_1,t) \phi^\ast_s(\x'_2,t) \phi_l(\x_2,t) \phi_{q}(\x_1,t)
\psi^\ast_{k'}(\y'_1,t) \psi_{q'}(\y_1,t), \nonumber \
\eeqn
where
\beq
\rho^{(AAB)}_{kk'slqq'}(t)=
\sum_{\vec J} C^\ast_{\vec J}(t) C^{\hat \rho^{(A)}_{kslq}\hat \rho^{(B)}_{k'q'}}_{\vec J}(t)
\eeq
are its matrix elements 
in the orbital basis.

\beqn\label{DNS_ABB}
& & \rho^{(ABB)}(\x_1,\y_1,\y_2|\x'_1,\y'_1,\y'_2;t) = \\
 & & = N_A N_B (N_B-1) \int \, d\x_2 \cdots d\x_{N_A} 
d\y_3 \cdots d\y_{N_B} d\z_1 \cdots d\z_{N_C} \times \nonumber \\
& & \times {\Psi^{(ABC)}}^\ast(\x'_1,\x_2,\ldots,\x_{N_A},\y'_1,\y'_2,\ldots,\y_{N_B},\z_1,\z_2,\ldots,\z_{N_C};t) 
\times \nonumber \\
& & \times \Psi^{(ABC)}(\x_1,\x_2,\ldots,\x_{N_A},\y_1,\y_2,\ldots,\y_{N_B},\z_1,\z_2,\ldots,\z_{N_C};t) = \nonumber \\
& & = \left<\Psi^{(ABC)}(t)\left| \left\{ 
\hat{\mathbf \Psi}_A^\dag(\x'_1)
\hat{\mathbf \Psi}_A(\x_1)
\hat{\mathbf \Psi}_B^\dag(\y'_1)
\hat{\mathbf \Psi}_B^\dag(\y'_2)
\hat{\mathbf \Psi}_B(\y_2)
\hat{\mathbf \Psi}_B(\y_1)
\right|\Psi^{(ABC)}(t)\right> \right\} = \nonumber \\
& & = \sum^{M_A,M_B}_{k,k',s',l',q,q'=1} 
\rho^{(ABB)}_{kk's'l'qq'}(t) 
\phi^\ast_{k}(\x'_1,t) 
\phi_{q}(\x_1,t) 
\psi^\ast_{k'}(\y'_1,t)
\psi^\ast_{s'}(\y'_2,t) 
\psi_{l'}(\y_2,t) 
\psi_{q'}(\y_1,t), \nonumber \
\eeqn
where
\beq
\rho^{(ABB)}_{kk's'l'qq'}(t)=
\sum_{\vec J} C^\ast_{\vec J}(t) C^{\hat \rho^{(A)}_{kq}\hat \rho^{(B)}_{k's'l'q'}}_{\vec J}(t)
\eeq
are its matrix elements in the orbital basis.

\beqn\label{DNS_AAC}
& & \rho^{(AAC)}(\x_1,\x_2,\z_1|\x'_1,\x'_2,\z'_1;t) = \\
 & & = N_A(N_A-1) N_C \int \, d\x_3 \cdots d\x_{N_A} 
d\y_1 \cdots d\y_{N_B} d\z_2 \cdots d\z_{N_C} \times \nonumber \\
& & \times {\Psi^{(ABC)}}^\ast(\x'_1,\x'_2,\ldots,\x_{N_A},\y_1,\y_2,\ldots,\y_{N_B},\z'_1,\z_2,\ldots,\z_{N_C};t) 
\times \nonumber \\
& & \times \Psi^{(ABC)}(\x_1,\x_2,\ldots,\x_{N_A},\y_1,\y_2,\ldots,\y_{N_B},\z_1,\z_2,\ldots,\z_{N_C};t) = \nonumber \\
& & = \left<\Psi^{(ABC)}(t)\left| \left\{ 
\hat{\mathbf \Psi}_A^\dag(\x'_1)
\hat{\mathbf \Psi}_A^\dag(\x'_2)
\hat{\mathbf \Psi}_A(\x_2)
\hat{\mathbf \Psi}_A(\x_1)
\hat{\mathbf \Psi}_C^\dag(\z'_1)
\hat{\mathbf \Psi}_C(\z_1)\right|\Psi^{(ABC)}(t)\right> \right\} = \nonumber \\
& & = \sum^{M_A,M_C}_{k,k'',s,l,q,q''=1} 
\rho^{(AAC)}_{kk''slqq''}(t) \phi^\ast_{k}(\x'_1,t) \phi^\ast_s(\x'_2,t) \phi_l(\x_2,t) \phi_{q}(\x_1,t)
\chi^\ast_{k''}(\z'_1,t) \chi_{q''}(\z_1,t), \nonumber \
\eeqn
where
\beq
\rho^{(AAC)}_{kk''slqq''}(t)=
\sum_{\vec J} C^\ast_{\vec J}(t) C^{\hat \rho^{(A)}_{kslq}\hat \rho^{(C)}_{k''q''}}_{\vec J}(t)
\eeq
are its matrix elements in the orbital basis. 

\beqn\label{DNS_ACC}
& & \rho^{(ACC)}(\x_1,\z_1,\z_2|\x'_1,\z'_1,\z'_2;t) = \\
 & & = N_A N_C (N_C-1) \int \, d\x_2 \cdots d\x_{N_A} 
d\y_1 \cdots d\y_{N_B} d\z_3 \cdots d\z_{N_C} \times \nonumber \\
& & \times {\Psi^{(ABC)}}^\ast(\x'_1,\x_2,\ldots,\x_{N_A},\y_1,\y_2,\ldots,\y_{N_B},\z'_1,\z'_2,\ldots,\z_{N_C};t) 
\times \nonumber \\
& & \times \Psi^{(ABC)}(\x_1,\x_2,\ldots,\x_{N_A},\y_1,\y_2,\ldots,\y_{N_B},\z_1,\z_2,\ldots,\z_{N_C};t) = \nonumber \\
& & = \left<\Psi^{(ABC)}(t)\left| \left\{ 
\hat{\mathbf \Psi}_A^\dag(\x'_1)
\hat{\mathbf \Psi}_A(\x_1)
\hat{\mathbf \Psi}_C^\dag(\z'_1)
\hat{\mathbf \Psi}_C^\dag(\z'_2)
\hat{\mathbf \Psi}_C(\z_2)
\hat{\mathbf \Psi}_C(\z_1)
\right|\Psi^{(ABC)}(t)\right> \right\} = \nonumber \\
& & = \sum^{M_A,M_C}_{k,k'',s'',l'',q,q''=1} 
\rho^{(ACC)}_{kk''s''l''qq''}(t) 
\phi^\ast_{k}(\x'_1,t) 
\phi_{q}(\x_1,t) 
\chi^\ast_{k''}(\z'_1,t)
\chi^\ast_{s''}(\z'_2,t) 
\chi_{l''}(\z_2,t) 
\chi_{q''}(\z_1,t), \nonumber \
\eeqn
where
\beq
\rho^{(ACC)}_{kk''s''l''qq''}(t)=
\sum_{\vec J} C^\ast_{\vec J}(t) C^{\hat \rho^{(A)}_{kq}\hat \rho^{(C)}_{k''s''l''q''}}_{\vec J}(t)
\eeq
are its matrix elements in the orbital basis. 

\beqn\label{DNS_BBC}
& & \rho^{(BBC)}(\y_1,\y_2,\z_1|\y'_1,\y'_2,\z'_1;t) = \\
 & & = N_B(N_B-1) N_C \int \, d\x_1 \cdots d\x_{N_A} 
d\y_3 \cdots d\y_{N_B} d\z_2 \cdots d\z_{N_C} \times \nonumber \\
& & \times {\Psi^{(ABC)}}^\ast(\x_1,\x_2,\ldots,\x_{N_A},\y'_1,\y'_2,\ldots,\y_{N_B},\z'_1,\z_2,\ldots,\z_{N_C};t) 
\times \nonumber \\
& & \times \Psi^{(ABC)}(\x_1,\x_2,\ldots,\x_{N_A},\y_1,\y_2,\ldots,\y_{N_B},\z_1,\z_2,\ldots,\z_{N_C};t) = \nonumber \\
& & = \left<\Psi^{(ABC)}(t)\left| \left\{ 
\hat{\mathbf \Psi}_B^\dag(\y'_1)
\hat{\mathbf \Psi}_B^\dag(\y'_2)
\hat{\mathbf \Psi}_B(\y_2)
\hat{\mathbf \Psi}_B(\y_1)
\hat{\mathbf \Psi}_C^\dag(\z'_1)
\hat{\mathbf \Psi}_C(\z_1)\right|\Psi^{(ABC)}(t)\right> \right\} = \nonumber \\
& & = \sum^{M_B,M_C}_{k',k'',s',l',q',q''=1} 
\rho^{(BBC)}_{k'k''s'l'q'q''}(t) \psi^\ast_{k'}(\y'_1,t) \psi^\ast_{s'}(\y'_2,t) \psi_{l'}(\y_2,t) \phi_{q'}(\y_1,t)
\chi^\ast_{k''}(\z'_1,t) \chi_{q''}(\z_1,t), \nonumber \
\eeqn
where
\beq
\rho^{(BBC)}_{k'k''s'l'q'q''}(t)=
\sum_{\vec J} C^\ast_{\vec J}(t) C^{\hat \rho^{(B)}_{k's'l'q'}\hat \rho^{(C)}_{k''q''}}_{\vec J}(t)
\eeq
are its matrix elements in the orbital basis. 

\beqn\label{DNS_BCC}
& & \rho^{(BCC)}(\y_1,\z_1,\z_2|\y'_1,\z'_1,\z'_2;t) = \\
 & & = N_B N_C (N_C-1) \int \, d\x_1 \cdots d\x_{N_A} 
d\y_2 \cdots d\y_{N_B} d\z_3 \cdots d\z_{N_C} \times \nonumber \\
& & \times {\Psi^{(ABC)}}^\ast(\x_1,\x_2,\ldots,\x_{N_A},\y'_1,\y_2,\ldots,\y_{N_B},\z'_1,\z'_2,\ldots,\z_{N_C};t) 
\times \nonumber \\
& & \times \Psi^{(ABC)}(\x_1,\x_2,\ldots,\x_{N_A},\y_1,\y_2,\ldots,\y_{N_B},\z_1,\z_2,\ldots,\z_{N_C};t) = \nonumber \\
& & = \left<\Psi^{(ABC)}(t)\left| \left\{ 
\hat{\mathbf \Psi}_B^\dag(\y'_1)
\hat{\mathbf \Psi}_B(\y_1)
\hat{\mathbf \Psi}_C^\dag(\z'_1)
\hat{\mathbf \Psi}_C^\dag(\z'_2)
\hat{\mathbf \Psi}_C(\z_2)
\hat{\mathbf \Psi}_C(\z_1)
\right|\Psi^{(ABC)}(t)\right> \right\} = \nonumber \\
& & = \sum^{M_B,M_C}_{k',k'',s'',l'',q',q''=1} 
\rho^{(BCC)}_{k'k''s''l''q'q''}(t) 
\psi^\ast_{k'}(\y'_1,t) 
\psi_{q'}(\y_1,t) 
\chi^\ast_{k''}(\z'_1,t)
\chi^\ast_{s''}(\z'_2,t) 
\chi_{l''}(\z_2,t) 
\chi_{q''}(\z_1,t), \nonumber \
\eeqn
where
\beq
\rho^{(BCC)}_{k'k''s''l''q'q''}(t)=
\sum_{\vec J} C^\ast_{\vec J}(t) C^{\hat \rho^{(B)}_{k'q'}\hat \rho^{(C)}_{k''s''l''q''}}_{\vec J}(t)
\eeq
are its matrix elements in the orbital basis.

Finally, there is a single reduced three-body density matrix
which is associated with
the three-body interaction of three
distinct particles.

\beqn\label{DNS_ABC}
& & \rho^{(ABC)}(\x_1,\y_1,\z_1|\x'_1,\y'_1,\z'_1;t) = \\
 & & = N_A N_B N_C \int \, d\x_2 \cdots d\x_{N_A} 
d\y_2 \cdots d\y_{N_B} d\z_2 \cdots d\z_{N_C} \times \nonumber \\
& & \times {\Psi^{(ABC)}}^\ast(\x'_1,\x_2,\ldots,\x_{N_A},\y'_1,\y_2,\ldots,\y_{N_B},\z'_1,\z_2,\ldots,\z_{N_C};t) 
\times \nonumber \\
& & \times \Psi^{(ABC)}(\x_1,\x_2,\ldots,\x_{N_A},\y_1,\y_2,\ldots,\y_{N_B},\z_1,\z_2,\ldots,\z_{N_C};t) = \nonumber \\
& & = \left<\Psi^{(ABC)}(t)\left| \left\{ 
\hat{\mathbf \Psi}_A^\dag(\x'_1)
\hat{\mathbf \Psi}_A(\x_1)
\hat{\mathbf \Psi}_B^\dag(\y'_1)
\hat{\mathbf \Psi}_B(\y_1)
\hat{\mathbf \Psi}_C^\dag(\z'_2)
\hat{\mathbf \Psi}_C(\z_1)
\right|\Psi^{(ABC)}(t)\right> \right\} = \nonumber \\
& & = \sum^{M_A,M_B,M_C}_{k,k',k'',q,q',q''=1} 
\rho^{(ABC)}_{kk'k''qq'q''}(t) 
\phi^\ast_{k}(\x'_1,t) 
\phi_{q}(\x_1,t) 
\psi^\ast_{k'}(\y'_1,t)
\psi_{q'}(\y_1,t), 
\chi^\ast_{k}(\y'_1,t) 
\chi_{q}(\y_1,t) 
\nonumber \
\eeqn
where
\beq
\rho^{(ABC)}_{kk'k''qq'q'}(t)=
\sum_{\vec J} C^\ast_{\vec J}(t) C^{\hat \rho^{(A)}_{kq}\hat \rho^{(B)}_{k'q'}\hat \rho^{(C)}_{k''q''}}_{\vec J}(t)
\eeq
are its matrix elements in the orbital basis.

%% \newpage

\section{Further details of the derivation of the equations-of-motion
for mixtures of three kinds of identical particles}\label{appendix_C}

The derivation of the equations-of-motion
for the orbitals (\ref{EOM_final_orbitals_3mix})
starts from expressing the expectation value
of $\hat H^{(ABC)}$ with respect to the
many-particle wave-function $\left|\Psi^{(ABC)}\right>$
in a form which depends explicitly
on the various integrals with respect to the orbitals.
Thus we have:
%%%%%%%%%%%%%% EXPECTATION VALUE OF H_ABC %%%%%%%%%%%%%%%%%%%%%%%%%
\beqn\label{expectation_ALL_orbitals}
& &\left<\Psi^{(ABC)}\left|\hat H^{(ABC)} 
- i\frac{\partial}{\partial t}\right|\Psi^{(ABC)}\right> = 
%%%%%%% A
\sum_{k,q=1}^{M_A} \rho^{(A)}_{kq} \left[ h^{(A)}_{kq} - 
\left\{i\frac{\partial}{\partial t}^{(A)}\right\}_{kq} \right] + \nonumber \\
&& + \frac{1}{2}\sum_{k,s,l,q=1}^{M_A} \rho^{(A)}_{kslq} W^{(A)}_{ksql} 
+ \frac{1}{6}\sum_{k,s,p,r,l,q=1}^{M_A} \rho^{(A)}_{ksprlq} U^{(A)}_{kspqlr} + \nonumber \\
%%%%%%% B
&& + \sum_{k',q'=1}^{M_B} \rho^{(B)}_{k'q'} \left[ h^{(B)}_{k'q'} - 
\left\{i\frac{\partial}{\partial t}^{(B)}\right\}_{k'q'} \right] + \nonumber \\
&& + \frac{1}{2}\sum_{k',s',l',q'=1}^{M_B} \rho^{(B)}_{k's'l'q'} W^{(B)}_{k's'q'l'} 
+ \frac{1}{6}\sum_{k',s',p',r',l',q'=1}^{M_B} \rho^{(B)}_{k's'p'r'l'q'} U^{(B)}_{k's'p'q'l'r'} + \nonumber \\
%%%%%%% C
&& + \sum_{k'',q''=1}^{M_C} \rho^{(C)}_{k''q''} \left[ h^{(C)}_{k''q''} - 
\left\{i\frac{\partial}{\partial t}^{(C)}\right\}_{k''q''} \right] + \nonumber \\
&& + \frac{1}{2}\sum_{k'',s'',l'',q''=1}^{M_C} \rho^{(C)}_{k''s''l''q''} W^{(C)}_{k''s''q''l''} + \nonumber \\
& & + \frac{1}{6}\sum_{k'',s'',p'',r'',l'',q''=1}^{M_C} \rho^{(C)}_{k''s''p''r''l''q''} U^{(C)}_{k''s''p''q''l''r''} 
+  \\
%%%%%%%% AB   AC   BC
& & + \sum_{k,k',q,q'=1}^{M_A,M_B} \rho^{(AB)}_{kk'qq'} W^{(AB)}_{kk'qq'}
    + \sum_{k,k'',q,q''=1}^{M_A,M_C} \rho^{(AC)}_{kk''qq''} W^{(AC)}_{kk''qq''} + \nonumber \\
& & + \sum_{k',k'',q',q''=1}^{M_B,M_C} \rho^{(BC)}_{k'k''q'q''} W^{(BC)}_{k'k''q'q''} + \nonumber \\
%%%%%%%% AAB ...
& & + \frac{1}{2} \sum_{k,k',s,q,q',l=1}^{M_A,M_B} \rho^{(AAB)}_{kk'slqq'} U^{(AAB)}_{kk'sqq'l} 
    + \frac{1}{2} \sum_{k,k',s',q,q',l'=1}^{M_A,M_B} \rho^{(ABB)}_{kk's'l'qq'} U^{(ABB)}_{kk's'qq'l'} + \nonumber \\
& & + \frac{1}{2} \sum_{k,k'',s,q,q'',l=1}^{M_A,M_C} \rho^{(AAC)}_{kk''slqq''} U^{(AAC)}_{kk''sqq''l}
    + \frac{1}{2} \sum_{k,k'',s'',q,q'',l''=1}^{M_A,M_C} \rho^{(ACC)}_{kk''s''l''qq''} U^{(ACC)}_{kk''s''qq''l''}
+ \nonumber \\
& & + \frac{1}{2} \sum_{k',k'',s',q',q'',l'=1}^{M_B,M_C} \rho^{(BBC)}_{k'k''s'l'q'q''} U^{(BBC)}_{k'k''s'q'q''l'} 
    + \frac{1}{2} \sum_{k',k'',s'',q',q'',l''=1}^{M_B,M_C} \rho^{(BCC)}_{k'k''s''l''q'q''} U^{(BCC)}_{k'k''s''q'q''l''} +
\nonumber \\
& & + \sum_{k,k',k'',q,q',q''=1}^{M_A,M_B,M_C} \rho^{(ABC)}_{kk'k''qq'q''} U^{(ABC)}_{kk'k''qq'q''} 
  - \sum_{\{\vec J\}}
 i C^\ast_{\vec J}(t) \dot C_{\vec J}(t). \nonumber \
\eeqn
The expectation values of the 
various density operators appearing in (\ref{expectation_ALL_orbitals})
have been prescribed in Appendix \ref{appendix_B}.

The matrix elements in (\ref{expectation_ALL_orbitals}) of the $A$ and correspondingly of the $B$ and $C$ 
single-species terms with respect to the orbitals have been discussed in section \ref{SEC2.1}, 
see Eq.~(\ref{matrix_elements}).
The matrix elements arising from two-body inter-species interactions
are listed for completeness below:
\beqn\label{MIX_matrix_elements_2B}
& & W^{(AB)}_{kk'qq'} = \int \!\! \int \phi_k^\ast(\x,t) \psi_{k'}^\ast(\y,t) \hat W^{(AB)}(\x,\y)
 \phi_q(\x,t) \psi_{q'}(\y,t) d\x d\y, \nonumber \\
& & W^{(AC)}_{kk''qq''} = \int \!\! \int \phi_k^\ast(\x,t) \chi_{k''}^\ast(\z,t) \hat W^{(AC)}(\x,\z)
 \phi_q(\x,t) \chi_{q''}(\z,t) d\x d\z, \nonumber \\
& & W^{(BC)}_{k'k''q'q''} = \int \!\! \int \psi_{k'}^\ast(\y,t) \chi_{k''}^\ast(\z,t) \hat W^{(BC)}(\y,\z)
 \psi_{q'}(\y,t) \chi_{q''}(\z,t) d\y d\z, \
\eeqn
and the matrix elements arising from 
three-body inter-species interactions read
as follows:
\beqn\label{MIX_matrix_elements_3B}
%%%%%%%% AAB ...
& &  U^{(AAB)}_{kk'sqq'l} = \nonumber \\ 
& & =  \int \!\! \int \!\! \int 
\phi_k^\ast(\x,t) \phi_s^\ast(\x',t) \psi_{k'}^\ast(\y,t) \hat U^{(AAB)}(\x,\x',\y)
 \phi_q(\x,t) \phi_l(\x',t) \psi_{q'}(\y,t) d\x d\x' d\y, \nonumber \\
%%%%%%% ABB 
& & U^{(ABB)}_{kk's'qq'l'} = \nonumber \\ 
& & =  \int \!\! \int \!\! \int 
\phi_k^\ast(\x,t) \psi_{k'}^\ast(\y,t) \psi_{s'}^\ast(\y',t) \hat U^{(ABB)}(\x,\y,\y')
 \phi_q(\x,t) \psi_{q'}(\y,t) \psi_{l'}(\y',t) d\x d\y d\y', \nonumber \\
%%%%%%% AAC
& &  U^{(AAC)}_{kk''sqq''l} = \nonumber \\ 
& & =  \int \!\! \int \!\! \int 
\phi_k^\ast(\x,t) \phi_s^\ast(\x',t) \chi_{k''}^\ast(\z,t) \hat U^{(AAC)}(\x,\x',\z)
 \phi_q(\x,t) \phi_l(\x',t) \chi_{q''}(\z,t) d\x d\x' d\z, \nonumber \\
%%%%%%% ACC
& &  U^{(ACC)}_{kk''s''qq''l''} = \nonumber \\ 
& & =  \int \!\! \int \!\! \int 
\phi_k^\ast(\x,t) \chi_{k''}^\ast(\z,t) \chi_{s''}^\ast(\z',t) \hat U^{(ACC)}(\x,\z,\z')
 \phi_q(\x,t) \chi_{q''}(\z,t) \chi_{l''}(\z',t) d\x d\z d\z', \nonumber \\
%%%%  BBC 
& & U^{(BBC)}_{k'k''s'q'q''l'} = \nonumber \\ 
& & =  \int \!\! \int \!\! \int 
\psi_{k'}^\ast(\y,t) \psi_{s'}^\ast(\y',t) \chi_{k''}^\ast(\z,t) \hat U^{(BBC)}(\y,\y',\z)
 \psi_{q'}(\y,t) \psi_{l'}(\y',t) \chi_{q''}(\z,t) d\y d\y' d\z, \nonumber \\
%%%%%% BCC
& & U^{(BCC)}_{k'k''s''q'q''l''} = \nonumber \\ 
& & =  \int \!\! \int \!\! \int 
\psi_{k'}^\ast(\y,t) \chi_{k''}^\ast(\z,t) \chi_{s''}^\ast(\z',t) \hat U^{(BCC)}(\y,\z,\z')
 \psi_{q'}(\y,t) \chi_{q''}(\z,t) \chi_{l''}(\z',t) d\y d\z d\z', \nonumber \\
%%%% ABC
& &  U^{(ABC)}_{kk'k''qq'q''} = \nonumber \\ 
& & =  \int \!\! \int \!\! \int 
\phi_{k}^\ast(\x,t)  \psi_{k'}^\ast(\y,t) \chi_{k''}^\ast(\z,t) \hat U^{(ABC)}(\x,\y,\z)
 \phi_{q}(\x,t) \psi_{q'}(\y,t) \chi_{q''}(\z,t) d\x d\y d\z. \nonumber \\
\eeqn

Performing the variation of the integrals (\ref{MIX_matrix_elements_2B})
with respect to the orbitals $\left\{\phi_k(\x,t)\right\}$,
$\left\{\psi_{k'}(\y,t)\right\}$
and
$\left\{\chi_{k''}(\z,t)\right\}$,
we find six types of inter-species 
one-body potentials
emerging 
from two-body interactions:
\beqn\label{all_local_2B_potentials}
& & \hat W^{(AB)}_{k'q'}(\x,t) = \int \psi_{k'}^\ast(\y,t)
\hat W^{(AB)}(\x,\y) \psi_{q'}(\y,t) d\y, \nonumber \\
& & \hat W^{(BA)}_{kq}(\y,t) = \int \phi_{k}^\ast(\x,t)
\hat W^{(AB)}(\x,\y) \phi_{q}(\x,t) d\x, \nonumber \\
& & \hat W^{(AC)}_{k''q''}(\x,t) = \int \chi_{k''}^\ast(\z,t)
\hat W^{(AC)}(\x,\z) \chi_{q''}(\z,t) d\z, \nonumber \\
& & \hat W^{(CA)}_{kq}(\z,t) = \int \phi_{k}^\ast(\x,t)
\hat W^{(AC)}(\x,\z) \phi_{q}(\x,t) d\x, \nonumber \\
& & \hat W^{(BC)}_{k''q''}(\y,t) = \int \chi_{k''}^\ast(\z,t)
\hat W^{(BC)}(\y,\z) \chi_{q''}(\z,t) d\z, \nonumber \\
& & \hat W^{(CB)}_{k'q'}(\z,t) = \int \psi_{k'}^\ast(\y,t)
\hat W^{(BC)}(\y,\z) \psi_{q'}(\y,t) d\y.
\eeqn
Making the variation of the integrals (\ref{MIX_matrix_elements_3B})
with respect to the orbitals,
we arrive at fifteen types
of inter-species one-body potentials 
resulting from three-body interactions:

\newpage

%%%%%%%%% from 3B forces
\beqn\label{all_local_3B_potentials}
& &  \hat U^{(AAB)}_{sk'lq'}(\x,t) =
 \int \!\! \int \phi_s^\ast(\x',t) \psi_{k'}^\ast(\y,t) \hat U^{(AAB)}(\x,\x',\y)
  \phi_l(\x',t) \psi_{q'}(\y,t) d\x' d\y, \nonumber \\
& &  \hat U^{(BAA)}_{ksql}(\y,t) = 
\int \!\! \int
\phi_k^\ast(\x,t) \phi_s^\ast(\x',t) \hat U^{(AAB)}(\x,\x',\y)
 \phi_q(\x,t) \phi_l(\x',t) d\x d\x', \nonumber \\
& & \hat U^{(ABB)}_{k's'q'l'}(\x,t)  
 =  \int \!\! \int  
  \psi_{k'}^\ast(\y,t) \psi_{s'}^\ast(\y',t) \hat U^{(ABB)}(\x,\y,\y')
   \psi_{q'}(\y,t) \psi_{l'}(\y',t) d\y d\y', \nonumber \\
& & \hat U^{(BAB)}_{ks'ql'}(\y,t) 
 =  \int \!\! \int  
\phi_k^\ast(\x,t) \psi_{s'}^\ast(\y',t) \hat U^{(ABB)}(\x,\y,\y')
 \phi_q(\x,t) \psi_{l'}(\y',t) d\x d\y', \nonumber \\
& & \hat U^{(AAC)}_{sk''lq''}(\x,t) 
  =  \int \!\! \int 
 \phi_s^\ast(\x',t) \chi_{k''}^\ast(\z,t) \hat U^{(AAC)}(\x,\x',\z)
  \phi_l(\x',t) \chi_{q''}(\z,t) d\x' d\z, \nonumber \\
& & \hat U^{(CAA)}_{ksql}(\z,t) 
  =  \int \!\! \int 
\phi_k^\ast(\x,t) \phi_s^\ast(\x',t) \hat U^{(AAC)}(\x,\x',\z)
 \phi_q(\x,t) \phi_l(\x',t) d\x d\x', \nonumber \\
& & \hat U^{(ACC)}_{k''s''q''l''}(\x,t)  
  =  \int \!\! \int 
  \chi_{k''}^\ast(\z,t) \chi_{s''}^\ast(\z',t) \hat U^{(ACC)}(\x,\z,\z')
   \chi_{q''}(\z,t) \chi_{l''}(\z',t) d\z d\z', \nonumber \\
& & \hat U^{(CAC)}_{ks''ql''}(\z,t)  
 =  \int \!\! \int  
\phi_k^\ast(\x,t) \chi_{s''}^\ast(\z',t) \hat U^{(ACC)}(\x,\z,\z')
 \phi_q(\x,t) \chi_{l''}(\z',t) d\x d\z', \nonumber \\
& & \hat U^{(BBC)}_{s'k''l'q''}(\y,t)  
  =  \int \!\! \int  
  \psi_{s'}^\ast(\y',t) \chi_{k''}^\ast(\z,t) \hat U^{(BBC)}(\y,\y',\z)
  \psi_{l'}(\y',t) \chi_{q''}(\z,t) d\y' d\z, \nonumber \\
& & \hat U^{(CBB)}_{k's'q'l'}(\z,t) 
=  \int \!\! \int 
\psi_{k'}^\ast(\y,t) \psi_{s'}^\ast(\y',t) \hat U^{(BBC)}(\y,\y',\z)
 \psi_{q'}(\y,t) \psi_{l'}(\y',t) d\y d\y', \nonumber \\
& & \hat U^{(BCC)}_{k''s''q''l''}(\y,t)
 =  \int \!\! \int
 \chi_{k''}^\ast(\z,t) \chi_{s''}^\ast(\z',t) \hat U^{(BCC)}(\y,\z,\z')
 \chi_{q''}(\z,t) \chi_{l''}(\z',t) d\z d\z', \nonumber \\
& & \hat U^{(CBC)}_{k's''q'l''}(\z,t) 
=  \int \!\! \int 
\psi_{k'}^\ast(\y,t) \chi_{s''}^\ast(\z',t) \hat U^{(BCC)}(\y,\z,\z')
 \psi_{q'}(\y,t) \chi_{l''}(\z',t) d\y d\z', \nonumber \\
& & \hat U^{(ABC)}_{k'k''q'q''}(\x,t) 
 =  \int \!\! \int 
  \psi_{k'}^\ast(\y,t) \chi_{k''}^\ast(\z,t) \hat U^{(ABC)}(\x,\y,\z)
  \psi_{q'}(\y,t) \chi_{q''}(\z,t) d\y d\z, \nonumber \\
& & \hat U^{(BAC)}_{kk''qq''}(\y,t) 
 =  \int \!\! \int  
\phi_{k}^\ast(\x,t) \chi_{k''}^\ast(\z,t) \hat U^{(ABC)}(\x,\y,\z)
 \phi_{q}(\x,t) \chi_{q''}(\z,t) d\x d\z, \nonumber \\
& & \hat U^{(CAB)}_{kk'qq'}(\z,t) 
 =  \int \!\! \int  
\phi_{k}^\ast(\x,t) \psi_{k'}^\ast(\y,t) \hat U^{(ABC)}(\x,\y,\z)
 \phi_{q}(\x,t) \psi_{q'}(\y,t) d\x d\y. \
\eeqn
All one-body potentials in (\ref{all_local_2B_potentials})
and (\ref{all_local_3B_potentials})
are local 
(for spin-independent interactions),
time-dependent potentials.

To arrive at the final form 
of the
equations-of-motion (\ref{EOM_final_orbitals_3mix}), 
we define the auxiliary one-body operators
for the $A$-species' 
particles:
\beqn\label{1B_oper_A}
& & \{\rho_2 \hat W\}^{(A)}_{kq} \equiv
 \sum^{M_A}_{s,l=1} \rho^{(A)}_{kslq} \hat W^{(A)}_{sl} +
\sum_{k',q'=1}^{M_B} \rho^{(AB)}_{kk'qq'} \hat W^{(AB)}_{k'q'} +
\sum_{k'',q''=1}^{M_C} \rho^{(AC)}_{kk''qq''} \hat W^{(AC)}_{k''q''}, \nonumber \\
& & \{\rho_3 \hat U\}^{(A)}_{kq} \equiv
 \frac{1}{2} \sum^{M_A}_{s,p,l,r=1} \rho^{(A)}_{ksprlq} \hat U^{(A)}_{splr} +
 \sum_{k',s,q',l=1}^{M_A,M_B} \rho^{(AAB)}_{kk'slqq'} \hat U^{(AAB)}_{sk'lq'} + \nonumber \\
& & + \sum_{k',s',q',l'=1}^{M_B} \rho^{(ABB)}_{kk's'l'qq'} \hat U^{(ABB)}_{k's'q'l'} + 
 \sum_{k'',s,q'',l=1}^{M_A,M_C} \rho^{(AAC)}_{kk''slqq''} \hat U^{(AAC)}_{sk''lq''} + \nonumber \\
& & + \sum_{k'',s'',q'',l''=1}^{M_C} \rho^{(ACC)}_{kk''s''l''qq''} \hat U^{(ACC)}_{k''s''q''l''} 
 + \sum_{k',k'',q',q''=1}^{M_B,M_C} \rho^{(ABC)}_{kk'k''qq'q''} \hat U^{(ABC)}_{k'k''q'q''}, \
\eeqn
for the $B$-species' particles:
\beqn\label{1B_oper_B}
& & \{\rho_2 \hat W\}^{(B)}_{k'q'} \equiv
 \sum^{M_B}_{s',l'=1} \rho^{(B)}_{k's'l'q'} \hat W^{(B)}_{s'l'} +
 \sum_{k,q=1}^{M_A} \rho^{(AB)}_{kk'qq'} \hat W^{(BA)}_{kq} +
 \sum_{k'',q''=1}^{M_C} \rho^{(BC)}_{k'k''q'q''} \hat W^{(BC)}_{k''q''}, \nonumber \\
& &  \{\rho_3 \hat U\}^{(B)}_{k'q'} \equiv
  \frac{1}{2}\sum_{s',p',r',l'=1}^{M_B} \rho^{(B)}_{k's'p'r'l'q'} \hat U^{(B)}_{s'p'l'r'} +
 \sum_{k,s,q,l=1}^{M_A} \rho^{(AAB)}_{kk'slqq'} \hat U^{(BAA)}_{ksql} + \nonumber \\
& & + \sum_{k,s',q,l'=1}^{M_A,M_B} \rho^{(ABB)}_{kk's'l'qq'} \hat U^{(BAB)}_{ks'ql'} +
 \sum_{k'',s',q'',l'=1}^{M_B,M_C} \rho^{(BBC)}_{k'k''s'l'q'q''} \hat U^{(BBC)}_{s'k''l'q''} + \nonumber \\
& & + \sum_{k'',s'',q'',l''=1}^{M_C} \rho^{(BCC)}_{k'k''s''l''q'q''} \hat U^{(BCC)}_{k''s''q''l''} +
\sum_{k,k'',q,q''=1}^{M_A,M_C} \rho^{(ABC)}_{kk'k''qq'q''} \hat U^{(BAC)}_{kk''qq''},
\eeqn
and for the $C$-species' particles:
\beqn\label{1B_oper_C}
& & \{\rho_2 \hat W\}^{(C)}_{k''q''} \equiv
 \sum^{M_C}_{s'',l''=1} \rho^{(C)}_{k''s''l''q''} \hat W^{(C)}_{s''l''} +
 \sum_{k,q=1}^{M_A} \rho^{(AC)}_{kk''qq''} \hat W^{(CA)}_{kq} +
 \sum_{k',q'=1}^{M_B} \rho^{(BC)}_{k'k''q'q''} \hat W^{(CB)}_{k'q'}, \nonumber \\
& &  \{\rho_3 \hat U\}^{(C)}_{k''q''} \equiv
  \frac{1}{2}\sum_{s'',p'',r'',l''=1}^{M_C} \rho^{(C)}_{k''s''p''r''l''q''} \hat U^{(C)}_{s''p''l''r''} +
\sum_{k,s,q,l=1}^{M_A} \rho^{(AAC)}_{kk''slqq''} \hat U^{(CAA)}_{ksql} + \nonumber \\
& & \sum_{k,s'',q,l''=1}^{M_A,M_C} \rho^{(ACC)}_{kk''s''l''qq''} U^{(CAC)}_{ks''ql''} +
 \sum_{k',s',q',l'=1}^{M_B} \rho^{(BBC)}_{k'k''s'l'q'q''} \hat U^{(CBB)}_{k's'q'l'} + \nonumber \\
& & \sum_{k',s'',q',l''=1}^{M_B,M_C} \rho^{(BCC)}_{k'k''s''l''q'q''} \hat U^{(CBC)}_{k's''q'l''} +
\sum_{k,k',q,q'=1}^{M_A,M_B} \rho^{(ABC)}_{kk'k''qq'q''} \hat U^{(CAB)}_{kk'qq'}.
\eeqn
These auxiliary one-body operators are constructed from
products of matrix elements of reduced density matrices of increasing order
(see Appendix \ref{appendix_B})
times the one-body potentials resulting from
interactions of the same order,
see Eqs.~(\ref{all_local_2B_potentials}) and (\ref{all_local_3B_potentials}).
The derivation of
the equations-of-motion
(\ref{EOM_final_orbitals_3mix})
is now fully completed.

\end{document}